# Koopman-Operator Spectral Decomposition for Nonlinear Motion Suppression in Dynamic Contrast-Enhanced MRI of the Head and Neck

Renjie He

Department of Radiation Oncology
The University of Texas MD Anderson Cancer Center, Houston, TX

## Abstract

**Background and Purpose.** Dynamic contrast-enhanced MRI (DCE-MRI) is an essential functional imaging modality for treatment planning, response assessment, and outcome prediction in head-and-neck squamous cell carcinoma (HNSCC). By tracking the temporal evolution of gadolinium-based contrast agent concentration, DCE-MRI provides voxel-wise pharmacokinetic parameters—including the volume transfer constant $K^{trans}$, the extravascular extracellular volume fraction $v_e$, and the plasma volume fraction $v_p$—that serve as quantitative biomarkers of tumor vascularity and permeability. However, the head-and-neck region presents uniquely challenging motion characteristics: involuntary swallowing events occur every 30–60 seconds, quasi-periodic respiratory motion introduces systematic spatial shifts, and persistent low-amplitude muscle contractions in the floor of mouth and tongue base add a stochastic component. These motion sources corrupt the observed signal in a fundamentally nonlinear manner. Unlike rigid-body motion commonly encountered in abdominal imaging, head-and-neck motion creates multiplicative coupling between the time-varying contrast enhancement function C(x,t) and the tissue deformation field u(x,t) through the spoiled gradient-recalled echo (SPGR) signal equation. This coupling renders conventional linear decomposition methods—including standard Dynamic Mode Decomposition (DMD) and principal component analysis—theoretically inadequate for clean enhancement–motion separation.

**Methods.** We develop a comprehensive motion suppression framework grounded in Koopman operator theory, the mathematical formalism introduced by Koopman and von Neumann in 1931–1932 that represents nonlinear dynamics as linear operations in an infinite-dimensional observable function space. Our approach encompasses three levels of Koopman approximation: (i) standard DMD as the baseline identity-observable case, (ii) Extended DMD (EDMD) with a physics-informed lifting dictionary comprising identity, element-wise quadratic, and spatial gradient observables designed to capture the specific nonlinearities of the SPGR signal model, and (iii) a deep Koopman autoencoder that learns the lifting function via neural network optimization with a four-component physics-informed loss function. We introduce time-course repetition as a novel technique to resolve the critical data-sufficiency bottleneck: by tiling the temporal series N times before decomposition, we preserve the Koopman eigenvalue spectrum while increasing the snapshot count m, thereby improving the numerical conditioning of the EDMD least-squares covariance estimation in the PCA compression step. The complete pipeline processes DCE-MRI slice-by-slice through block decomposition, optional repetition, Koopman lifting, PCA compression, optional robust PCA for transient separation, exact DMD with spectral mode classification at the enhancement–motion frequency boundary, and block reassembly with raised-cosine blending.

**Results.** On a physics-based 128×128 head-and-neck phantom with tunable nonlinear coupling parameter γ, we demonstrate that at baseline (no repetition), Koopman EDMD provides only marginal improvement over standard DMD (MSE improvement ratio 1.796 vs. 1.780, SSIM 0.9819 vs. 0.9817). This confirms the $p/m$ ratio problem: with p≈700 lifted dimensions and only

m=50 snapshots, the EDMD covariance estimation is poorly conditioned. With 3× time-course repetition, EDMD achieves ratio 1.953 and SSIM 0.9835, representing the best overall performance—a 3.3% relative improvement over standard DMD. The coupling sweep confirms that the EDMD advantage grows monotonically with nonlinear coupling strength. Preliminary application to clinical HNC DCE-MRI data (5 axial slices, 55 temporal frames, 256×256 pixels) demonstrates successful motion–enhancement separation, with clear spectral separation of slowly varying pharmacokinetic enhancement modes from higher-frequency respiratory motion modes at the $f_c$ = 0.08 Hz boundary.

**Conclusions.** Koopman operator theory provides a principled, mathematically grounded framework for nonlinear motion suppression in DCE-MRI. The key practical contribution is time-course repetition, which resolves the data-sufficiency bottleneck that has limited Koopman lifting in medical imaging. All source code, clinical data, experimental notebooks, video demonstrations, and the complete manuscript are released as an open-source repository for full reproducibility.

# 1. Introduction

## 1.1. Clinical Context: DCE-MRI in Head-and-Neck Cancer

Head-and-neck squamous cell carcinoma (HNSCC) is the sixth most common cancer worldwide, with over 900,000 new cases annually. Treatment typically involves concurrent chemoradiotherapy, where the radiation dose is sculpted to the tumor volume using image-guided techniques. DCE-MRI has emerged as a critical tool in this treatment paradigm, providing functional information that complements anatomical $T_2$-weighted and diffusion-weighted imaging. The pharmacokinetic parameters derived from DCE-MRI—particularly $K^{trans}$ (the volume transfer constant reflecting vascular permeability) and $v_e$ (the extravascular extracellular volume fraction)—have been shown to predict treatment response, differentiate radiation necrosis from recurrence, and guide adaptive replanning strategies.

The standard DCE-MRI acquisition protocol for HNSCC involves a pre-contrast $T_1$ mapping sequence followed by a dynamic SPGR or fast spoiled gradient echo (FSPGR) series during and after intravenous injection of a gadolinium chelate (typically gadobutrol or gadopentetate dimeglumine at 0.1 mmol/kg). The temporal resolution is typically 3–5 seconds per volume, with 40–60 time points acquired over 2–4 minutes. Spatial resolution is 1–2 mm in-plane with 3–5 mm slice thickness, covering the primary tumor and draining lymph node stations from skull base to clavicles. The arterial input function (AIF) is typically measured from the internal carotid artery or a branch of the external carotid system, though AIF measurement itself is challenging in the head and neck due to the small caliber of the feeding vessels and susceptibility artifacts from the skull base and dental hardware.

Pharmacokinetic analysis proceeds by fitting the extended Tofts model to the measured tissue concentration-time curves. The model parameters—$K^{trans}$ (volume transfer constant, reflecting capillary permeability), $v_e$ (extravascular extracellular volume fraction), and $v_p$ (plasma volume fraction)—have established clinical significance: $K^{trans}$ is elevated in aggressive, well-vascularized tumors and decreases in response to effective treatment; $v_e$ reflects the extracellular space available for contrast agent distribution; and $v_p$ provides a measure of blood volume that can indicate treatment-induced vascular normalization. Changes in these parameters during the first 2–3 weeks of chemoradiotherapy have been shown to predict ultimate locoregional control, making early DCE-MRI a potential tool for adaptive treatment modification.

The clinical utility of DCE-MRI depends critically on the accuracy of pharmacokinetic model fitting, which in turn requires clean, motion-free enhancement time curves. The Tofts model, the most widely used pharmacokinetic framework, relates the tissue contrast agent concentration $C_t(t)$ to the arterial input function $C_p(t)$ through a convolution integral involving $K^{trans}$ and $v_e$. Even small perturbations in the measured signal—such as those introduced by sub-voxel motion—can propagate into large errors in the estimated pharmacokinetic parameters, particularly when the fitting involves nonlinear least-squares optimization over noisy data.

The head-and-neck region presents a uniquely challenging motion environment for DCE-MRI. Unlike the abdomen, where respiratory motion is primarily a rigid or affine displacement that can be addressed by navigator-based techniques or breath-holding, head-and-neck motion involves multiple independent and interacting sources. Swallowing is the dominant corruption source: it produces sudden, large-amplitude, spatially heterogeneous deformations in the floor of mouth, tongue base, and pharyngeal walls, occurring involuntarily every 30–60 seconds regardless of patient instruction. Respiratory motion adds a quasi-periodic component at approximately 0.15–0.30 Hz (12–18 breaths per minute), primarily visible as anterior-posterior displacement of the mandible and soft palate. Additionally, patients exhibit persistent low-amplitude muscle contractions (fasciculations) in the tongue and floor-of-mouth musculature, producing a stochastic, broadband motion component that is spatially concentrated at tissue interfaces.

## 1.2. The Nonlinear Motion–Enhancement Coupling Problem

The fundamental challenge in DCE-MRI motion correction for the head and neck is not the motion itself, but its nonlinear coupling with the contrast enhancement signal. In the SPGR sequence used for most DCE-MRI protocols, the observed signal intensity S depends on both the local $T_1$ relaxation time (which changes with contrast agent concentration) and the local tissue position (which changes with motion):

$$S(x,t) = M_0(x+u(x,t)) \cdot \sin\alpha \cdot [1 - \exp(-TR/T_1(x+u(x,t),t))] / [1 - \cos\alpha \cdot \exp(-TR/T_1(x+u(x,t),t))] \tag{1}$$

where $M_0$ is the equilibrium magnetization (a tissue property), α is the flip angle, TR is the repetition time, $T_1$(x,t) depends on the contrast agent concentration through the relaxivity relation $1/T_1(t) = 1/T_{10} + r_1 \cdot C(t)$, and u(x,t) is the displacement field. The critical observation is that this signal equation is nonlinear in both the contrast concentration C and the displacement u, and—most importantly—C and u appear in a multiplicatively coupled form. When tissue at position x moves to position x+u while simultaneously experiencing contrast enhancement, the resulting signal change is not simply the sum of a motion component and an enhancement component. Rather, the motion modulates the enhancement signal (and vice versa), creating cross-terms that cannot be separated by any linear filter.

This multiplicative coupling has concrete consequences: a linear decomposition method (such as standard DMD or PCA) will misattribute some of the cross-term energy to either the enhancement or motion component, depending on the relative amplitudes and frequencies. In practice, this manifests as residual motion texture in the ‘enhancement-only’ reconstruction and/or phantom enhancement signal in the ‘motion-only’ component. The severity of this leakage increases with both the motion amplitude and the contrast enhancement level, making it worst precisely where clinical accuracy matters most—in highly enhancing tumor tissue during peak contrast wash-in.

## 1.3. Prior Work: Motion Correction Approaches for DCE-MRI

Existing approaches to DCE-MRI motion correction fall into two broad categories. Registration-based methods attempt to align each temporal frame to a reference image through deformable image registration (DIR). While DIR has been successful in the abdomen and chest, it faces fundamental difficulties in the head and neck: (1) the moving tissue is often blurred and noisy, making intensity-based similarity metrics unreliable; (2) swallowing events produce non-diffeomorphic deformations (tissue folding) that violate the smoothness assumptions of most DIR algorithms; (3) the contrast enhancement itself changes the image appearance over time, confounding the registration cost function; and (4) the computational burden of volumetric deformable registration on 40–60 temporal frames is substantial.

Data-driven decomposition methods—including PCA, independent component analysis (ICA), and DMD—take a fundamentally different approach: rather than attempting to undo the motion, they decompose the spatiotemporal data into components that can be classified as ‘enhancement’ or ‘motion’ based on their temporal characteristics. DMD is particularly natural for this task because it directly yields temporal frequencies for each spatial mode, enabling physics-based classification. However, standard DMD assumes linear dynamics, which is precisely the assumption that fails when contrast and motion are multiplicatively coupled.

A separate line of work has explored model-driven approaches to DCE-MRI motion correction. Pharmacokinetic-aware registration jointly estimates motion and enhancement parameters by incorporating the Tofts model as a constraint in the registration cost function. While elegant, these methods are computationally expensive (requiring iterative nonlinear optimization over both deformation fields and pharmacokinetic maps) and sensitive to initialization. Neural network-based approaches, including convolutional neural networks trained on synthetic motion-corrupted data, have shown promise for fast inference but require large labeled training datasets that are difficult to obtain for head-and-neck anatomy.

The connection between DMD and video surveillance (background/foreground separation) has been explored by Erichson, Brunton, and Kutz through randomized DMD (rDMD). Their formulation decomposes a video matrix into a low-rank background and sparse foreground using DMD modes classified by their temporal dynamics. Our approach extends this concept from surveillance video to medical imaging, with Koopman lifting providing the nonlinearity handling that is unnecessary for static-background surveillance but essential for contrast-enhanced dynamic imaging. The multiresolution DMD (mrDMD) framework of Kutz et al. provides a hierarchical decomposition at multiple temporal scales, analogous to wavelet analysis. This approach handles non-stationary dynamics by recursively separating slow and fast modes. While mrDMD addresses temporal non-stationarity, our Koopman lifting addresses spatial nonlinearity—the two approaches are complementary and could potentially be combined.

Proctor, Brunton, and Kutz extended Koopman theory to systems with external inputs (DMDc), which is relevant when the contrast injection can be modeled as a known input signal. Korda and Mezic developed Koopman-based model predictive control (MPC), demonstrating that EDMD with radial basis function dictionaries can effectively linearize nonlinear control systems. Their least-squares formulation for the Koopman approximation matrix K (Equation 19–20 in our notation) is the mathematical foundation we adopt for DCE-MRI EDMD. The connection between Koopman eigenfunctions and system observability has been explored by several groups, establishing that a 'good' Koopman dictionary must approximate the system's eigenfunctions—a theoretical requirement that motivates our physics-informed design.

## 1.4. Koopman Operator Theory: A Path to Nonlinear Spectral Decomposition

Koopman operator theory provides a mathematically rigorous resolution to the linearity limitation. The key insight, formalized by Koopman in 1931 and elaborated by von Neumann in 1932, is that any nonlinear dynamical system can be represented as a linear system in an appropriately chosen function space. Specifically, the Koopman operator K acts on observable functions of the state (rather than on the state directly), and this action is always linear, regardless of the nonlinearity of the underlying state dynamics. The trade-off is dimensionality: the Koopman operator is generally infinite-dimensional, and practical implementations must approximate it with a finite-dimensional matrix. The quality of this approximation depends on the choice of observable functions (the 'lifting dictionary') and the amount of available data.

The past decade has seen an explosion of interest in computational Koopman methods, driven by Williams, Kevrekidis, and Rowley's Extended DMD (2015), the kernel DMD formulation of Kawahara (2016), and the deep learning approaches of Lusch, Brunton, and Kutz (2018) and Takeishi, Kawahara, and Yairi (2017). These methods have been applied successfully to fluid dynamics, robotics, power systems, and molecular dynamics. However, to our knowledge, the present work represents the first systematic application of the full Koopman operator framework—from classical EDMD through deep Koopman autoencoders—to medical imaging, and the first to identify and resolve the data-sufficiency bottleneck through time-course repetition.

## 1.5. Contributions of This Work

This paper makes the following contributions:

(1) We develop a physics-informed Koopman lifting dictionary specifically designed for the SPGR signal model, with quadratic observables targeting $T_1$-relaxation nonlinearity and gradient observables targeting motion-induced boundary corruption.

(2) We identify the $p/m$ ratio bottleneck that prevents Koopman lifting from improving over standard DMD in typical DCE-MRI acquisitions, and introduce time-course repetition as a simple technique that improves the conditioning of the EDMD covariance estimation.

(3) We implement and evaluate three levels of Koopman approximation—standard DMD, EDMD with hand-crafted dictionary, and deep Koopman autoencoder—on a physics-based phantom with tunable nonlinear coupling.

(4) We demonstrate preliminary clinical feasibility on HNC DCE-MRI data, with real computational results including spectral analysis, lifting visualization, and full-slice decomposition.

(5) We release all code, data, notebooks, and video demonstrations as an open-source repository for full reproducibility.

## 2. Mathematical Foundations

This section develops the mathematical framework from first principles, progressing from the abstract Koopman operator definition through its spectral decomposition, the DMD and EDMD algorithms for finite-dimensional approximation, and the deep learning extension. We aim for a self-contained treatment that is accessible to the medical imaging audience while maintaining mathematical rigor.

### 2.1. Dynamical Systems and the Koopman Operator

Consider a discrete-time dynamical system on a state space $M \subseteq \mathbb{R}^n$:

$$x_{k+1} = F(x_k), \quad x \in M \subseteq \mathbb{R}^n \tag{2}$$

where F: M → M is the evolution operator (possibly nonlinear). For a DCE-MRI time series, the state $x_k$ is the vectorized image at time step k, and F encapsulates all physical processes—contrast enhancement, motion, noise—that transform one frame into the next. The evolution operator F acts on state space: it takes a point in $\mathbb{R}^n$ and produces another point in $\mathbb{R}^n$.

The Koopman operator takes a fundamentally different perspective. Rather than acting on states, it acts on scalar-valued observable functions $\psi: M \to \mathbb{C}$. The Koopman operator K is defined as:

$$(K\psi)(x) = \psi(F(x)) = \psi \circ F(x) \tag{3}$$

In words: applying K to an observable ψ produces a new observable Kψ, whose value at any state x equals the value of the original observable ψ evaluated at the evolved state F(x). The fundamental and remarkable property is that K is always a linear operator, even when F is nonlinear. This follows immediately from the definition: for any observables $\psi_1$, $\psi_2$ and scalars α, β, we have $K(\alpha\psi_1 + \beta\psi_2)(x) = (\alpha\psi_1 + \beta\psi_2)(F(x)) = \alpha\psi_1(F(x)) + \beta\psi_2(F(x)) = \alpha(K\psi_1)(x) + \beta(K\psi_2)(x)$. The linearity holds because K acts on the function, not on the argument.

The trade-off is dimensionality: while F acts on the finite-dimensional state space $\mathbb{R}^n$, K acts on an infinite-dimensional function space. We have traded finite-dimensional nonlinearity for infinite-dimensional linearity. The practical question becomes: can we find a finite-dimensional subspace of the observable space where K is well-approximated by a matrix?

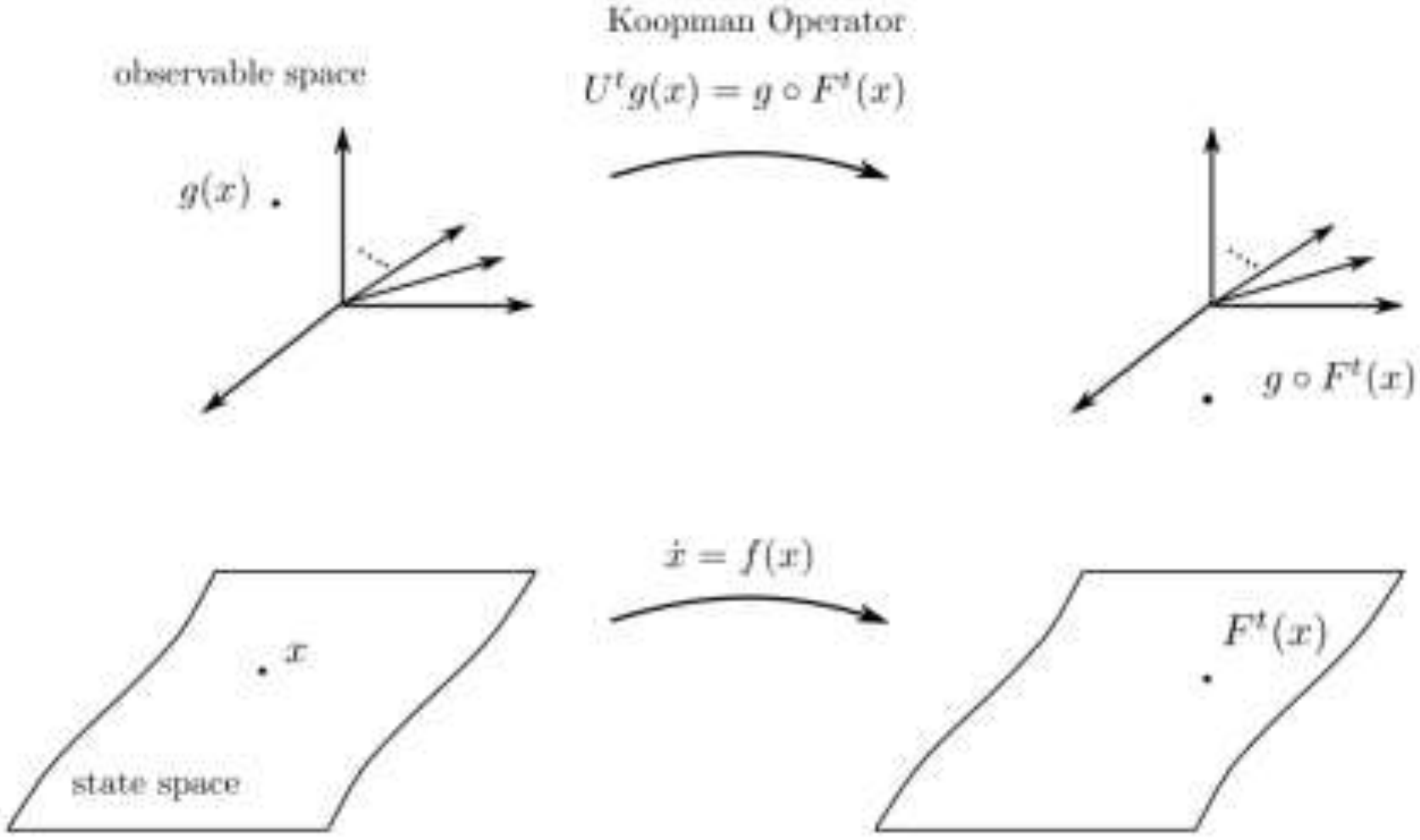

*Figure 1. The Koopman operator commutative diagram. Bottom: the nonlinear evolution operator F maps state x to $F^t(x)$ in the finite-dimensional state space M. The dynamics here may be highly nonlinear. Top: the Koopman operator $U^t$ propagates observables g(x) to $g(F^t(x))$ in the infinite-dimensional observable space, where the dynamics are guaranteed to be linear. The vertical arrows represent the lifting map g (upward: state → observables) and the projection/mode reconstruction (downward: observables → state). The commutativity of the diagram expresses the defining property of the Koopman operator: lifting then propagating is equivalent to propagating then lifting.*

## 2.2. Koopman Spectral Decomposition

Like any linear operator, K admits a spectral decomposition. A Koopman eigenfunction $\varphi_k$ is an observable that satisfies:

$$K\varphi_k = \mu_k \varphi_k \quad (4)$$

where $\mu_k \in \mathbb{C}$ is the corresponding Koopman eigenvalue. The eigenfunction $\varphi_k$ evolves in a particularly simple way under the dynamics: $\varphi_k(F(x)) = \mu_k \varphi_k(x)$. That is, $\varphi_k$ simply gets multiplied by a complex scalar at each time step. After t steps:

$$\varphi_k(F^t(x)) = \mu_k{}^t \varphi_k(x) \quad (5)$$

The eigenvalue $\mu_k$ encodes the complete temporal dynamics of its eigenfunction: the magnitude $|\mu_k|$ determines growth ($|\mu_k|>1$), decay ($|\mu_k|<1$), or neutrality ($|\mu_k|=1$), while the argument $\arg(\mu_k)/\Delta t$ gives the oscillation frequency in radians per second. Converting to the continuous-time representation via $\omega_k = \ln(\mu_k)/\Delta t$, the real part $\mathrm{Re}(\omega_k)$ is the growth rate and the imaginary part $\mathrm{Im}(\omega_k)/(2\pi)$ is the frequency in Hz.

If the Koopman operator has a complete set of eigenfunctions (which requires a purely discrete spectrum), any full-state observable g(x) = x can be expanded as a linear combination:

$$g(x) = x = \Sigma_k \, \xi_k \, \varphi_k(x) \quad (6)$$

where $\xi_k \in \mathbb{R}^n$ are the Koopman modes—the vector-valued coefficients that project each eigenfunction back to the physical state space. Substituting (5) into (6) gives the fundamental prediction equation:

$$F^t(x) = \Sigma_k \, \xi_k \, \mu_k{}^t \varphi_k(x) \tag{7}$$

This is an extraordinarily powerful result. It says that the future state of an arbitrarily nonlinear system can be computed by: (1) evaluating the eigenfunctions at the current state, (2) multiplying each by the appropriate eigenvalue raised to the t-th power, and (3) weighting by the modes to reconstruct the physical state. The entire nonlinear dynamics has been reduced to complex exponentials—precisely the structure that enables spectral classification for DCE-MRI motion suppression.

For DCE-MRI, this decomposition has a direct physical interpretation. Enhancement dynamics are slowly varying (pharmacokinetic wash-in/wash-out over minutes), corresponding to low-frequency eigenvalues near the positive real axis in the complex plane. Respiratory motion oscillates at 0.15–0.30 Hz, corresponding to eigenvalues with significant imaginary parts. Swallowing transients appear as eigenvalues with large growth/decay rates (far from the unit circle). By classifying eigenvalues and reconstructing from only the low-frequency enhancement modes, we achieve the desired motion–enhancement separation.

## 2.3. Dynamic Mode Decomposition (DMD)

DMD, introduced by Schmid in 2010, provides a practical algorithm for approximating the Koopman operator from snapshot data. Consider a time-ordered sequence of n observations:

$$V = \{v_1, v_2, v_3, \ldots, v_n\} \tag{8}$$

We seek a linear operator A such that $v_{k+1} \approx Av_k$ for all k. Partitioning the data into input and output matrices:

$$X = [v_1, v_2, \ldots, v_{n-1}], \quad Y = [v_2, v_3, \ldots, v_n] \tag{9}$$

the problem $Y \approx AX$ can be solved as $A = YX^+$ (pseudoinverse). However, for high-dimensional data (images with m pixels), forming the m×m matrix A explicitly is prohibitive. DMD circumvents this by projecting onto the dominant singular vectors of X.

### 2.3.1. The SVD Step

Compute the truncated SVD of X:

$$X = U\Sigma W^H \tag{10}$$

where $U \in \mathbb{R}^{m\text{x}r}$ contains the left singular vectors (spatial POD modes), $\Sigma \in \mathbb{R}^{r\text{x}r}$ contains the singular values in decreasing order, and $W \in \mathbb{R}^{n-1\text{x}r}$ contains the right singular vectors (temporal coefficients). The truncation rank r is chosen either as a fixed parameter or via a threshold on the singular value ratio $\sigma_i/\sigma_1$.

#### 2.3.2. The Reduced Operator

Project A onto the subspace spanned by U to obtain the r×r reduced operator:

$$\bar{A} = U^H\ A\ U = U^H\ Y\ W\ \Sigma^{-1} \tag{11}$$

The key insight is that Ā has the same eigenvalues as the best rank-r approximation of A (they are similar matrices in the subspace spanned by U). Computing Ā requires only r×r matrix operations, regardless of the ambient dimension m.

#### 2.3.3. Eigendecomposition and Mode Recovery

Compute the eigendecomposition of Ā:

$$\bar{A}\ \xi = \mu\ \xi \tag{12}$$

where μ are the DMD eigenvalues (approximating Koopman eigenvalues) and ξ the eigenvectors. The exact DMD modes—the spatial patterns associated with each temporal frequency—are recovered by projecting back to the full-dimensional space:

$$\Phi = Y\ W\ \Sigma^{-1}\ \xi \tag{13}$$

#### 2.3.4. Amplitudes and Prediction

The amplitude of each mode is determined by the initial condition via a least-squares fit:

$$b = \Phi^{+}\ v_1 \tag{14}$$

where $\Phi^{+}$ is the pseudoinverse of the mode matrix. The full reconstruction and t-step prediction are then:

$$v_k = \Sigma_j\ \varphi_j\ b_j\ \mu_j^{\ k} \ \ \textit{or equivalently} \ \ v_k = \Phi\ diag(b)\ \Lambda^k \tag{15}$$

where Λ is the diagonal matrix of eigenvalues. This is the discrete analog of Equation (7) for the standard DMD case.

#### 2.3.5. Worked Example: DMD on a Two-Mode System

To build intuition, consider a simple example: a 1D signal composed of two modes with frequencies $f_1 = 0.02$ Hz (slow enhancement) and $f_2 = 0.20$ Hz (respiratory motion), sampled at $\Delta t = 3.5$ seconds. The continuous-time eigenvalues are $\omega_1 = 2\pi \cdot 0.02i$ and $\omega_2 = 2\pi \cdot 0.20i$. The discrete-time DMD eigenvalues are $\mu_1 = \exp(\omega_1 \Delta t) = \exp(0.44i) \approx 0.905 + 0.426i$ and $\mu_2 = \exp(\omega_2 \Delta t) = \exp(4.40i) \approx -0.308 - 0.951i$. In the complex plane, $\mu_1$ sits near the top of the unit circle (small angle from the real axis, indicating slow oscillation), while $\mu_2$ sits in the third quadrant (large angle, rapid oscillation). The frequency cutoff $f_c = 0.08$ Hz correctly classifies $\mu_1$ as enhancement and $\mu_2$ as motion. Reconstructing from $\mu_1$ alone yields the clean enhancement signal; reconstructing from $\mu_2$ alone yields the motion component.

This example also illustrates why the SVD truncation rank r matters. With r = 1, only the dominant mode is captured (whichever has larger amplitude). With r = 2, both modes are resolved and can be classified independently. With r >> 2 on noisy data, noise modes enter the decomposition and contaminate the reconstruction. The optimal $r \approx 8$–12 for DCE-MRI reflects the typical number of physically meaningful spatiotemporal patterns: 2–3 enhancement modes (wash-in, wash-out, recirculation), 2–4 respiratory modes (fundamental + harmonics), and 1–2 baseline/drift modes.

DMD is mathematically equivalent to computing the Koopman decomposition under the identity observable $g(x) = x$. This is its fundamental limitation: when the underlying dynamics F are nonlinear, the identity observable does not span a Koopman-invariant subspace, and the linear approximation error in A can be substantial. The next section addresses this limitation.

## 2.4. Extended Dynamic Mode Decomposition (EDMD)

EDMD, introduced by Williams, Kevrekidis, and Rowley in 2015, generalizes DMD by replacing the identity observable with a richer dictionary of nonlinear functions. The idea is to lift the state into a higher-dimensional space where the Koopman operator is better approximated by a finite matrix.

### 2.4.1. The Dictionary of Observables

Define a set of K scalar-valued basis functions—the dictionary of observables:

$$D = \{\psi_1, \psi_2, \ldots, \psi_k\} \quad (16)$$

and construct the vector-valued lifting function:

$$\Psi(x) = [\psi_1(x), \psi_2(x), \ldots, \psi_k(x)]^T \in \mathbb{R}^K \quad (17)$$

The choice of dictionary is the central design decision in EDMD. The dictionary must be rich enough to approximate the Koopman eigenfunctions as linear combinations: $\varphi_j \approx \Sigma_k\, a_{jk}\psi_k$. If the dictionary spans (or closely approximates) a Koopman-invariant subspace, the finite-dimensional approximation will be accurate. Conversely, a poorly chosen dictionary will yield an EDMD that is no better—or even worse—than standard DMD, because the additional dimensions dilute the useful information without contributing Koopman-invariant content.

### 2.4.2. The Least-Squares Koopman Approximation

Given training data pairs $(x_m, y_m)$ with $y_m = F(x_m)$, EDMD seeks the K×K matrix K that minimizes the residual in the lifted space:

$$J = (1/2M)\, \Sigma_m \, \|\Psi(y_m) - K\Psi(x_m)\|^2 \quad (18)$$

This is a standard least-squares problem with the closed-form solution:

$$K = G^+ A \quad (19)$$

where the K×K matrices G and A are:

$$G = (1/M)\, \Sigma_m\, \Psi(x_m)^* \Psi(x_m), \quad A = (1/M)\, \Sigma_m\, \Psi(x_m)^* \Psi(y_m) \quad (20)$$

The matrix K is the finite-dimensional Koopman approximation. Its eigenvalues approximate the Koopman eigenvalues, its right eigenvectors $\xi_j$ yield approximate eigenfunctions via $\varphi_j = \Psi\xi_j$, and its left eigenvectors yield the approximate Koopman modes.

### 2.4.3. Recovery of the Three Koopman Elements

From the eigendecomposition of K:

Koopman eigenvalues: the eigenvalues $\mu_j$ of K directly approximate the Koopman eigenvalues.

Koopman eigenfunctions: given the right eigenvectors $\xi_j$ of K, the approximate eigenfunction is $\varphi_j(x) = \Psi(x)^T \xi_j$. This is a scalar function of the state x, evaluated by lifting x into the dictionary and projecting onto the eigenvector.

Koopman modes: the left eigenvectors $w_j$ of K (satisfying $w_j^H \xi_j = 1$) contribute to the mode reconstruction $v = (W^*B)^T$ where B encodes the relationship between the dictionary and the state variables.

The fundamental difference between DMD and EDMD is now clear. DMD uses the identity dictionary $\Psi(x) = x$, which gives K = A (the best-fit linear operator in the original state space). EDMD uses a nonlinear dictionary that can capture dynamics invisible to the identity observable. However, this comes at a cost: the lifted dimension K is larger than the original dimension n, requiring more data (larger m) to estimate K reliably—the $p/m$ ratio problem that motivates our time-course repetition approach.

## 2.5. Deep Koopman Autoencoders

Rather than hand-crafting the dictionary, deep Koopman methods use neural networks to learn the lifting function and its inverse. The architecture consists of three components: an encoder network $g: \mathbb{R}^n \to \mathbb{R}^K$ that learns the Koopman lifting, a linear propagation layer $K \in \mathbb{R}^{K \times K}$ that approximates the Koopman operator in the latent space, and a decoder network $h: \mathbb{R}^K \to \mathbb{R}^n$ that learns the inverse lifting (mode reconstruction).

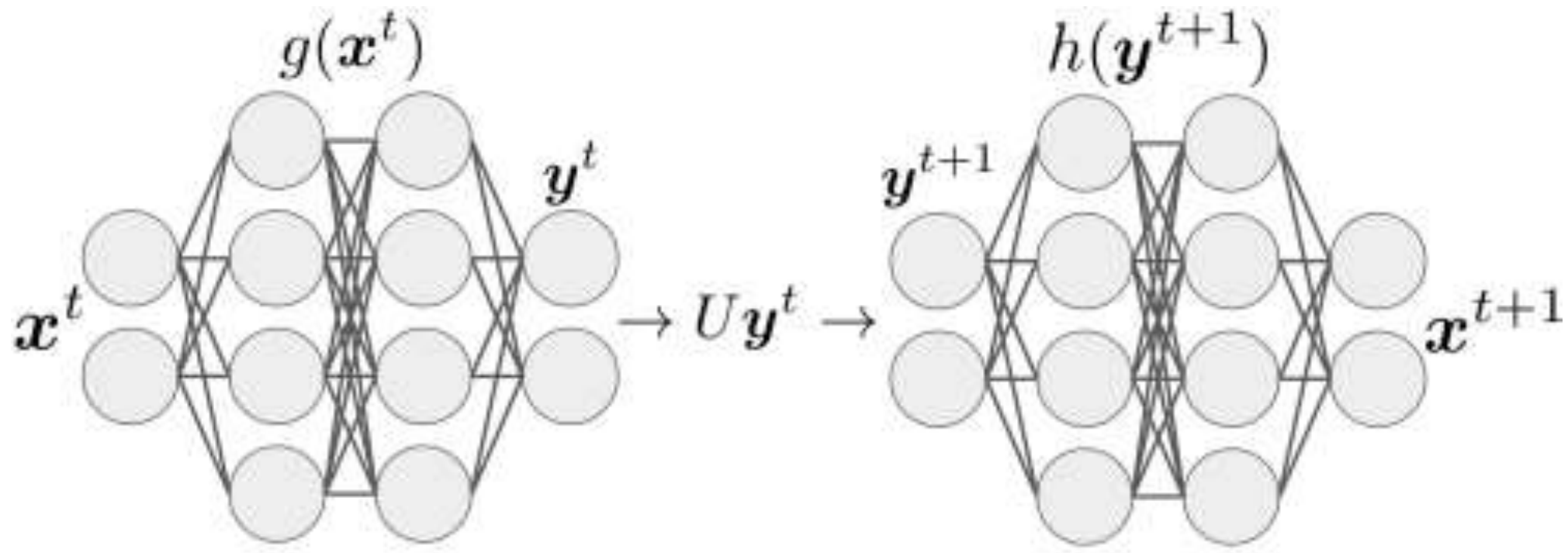

*Figure 2. Deep Koopman autoencoder architecture. Left: encoder g(·) maps state $x^t$ to latent embedding $y^t = g(x^t)$. The encoder is a multi-layer neural network that learns the Koopman observable dictionary. Center: the Koopman matrix U propagates the latent state linearly: $y^{t+1} = Uy^t$. Unlike fully data-driven approaches, U is computed analytically via the EDMD least-squares solution (Eq. 19–20) on the encoded snapshots, not learned by backpropagation. This makes the latent dynamics provably linear (within reconstruction error) and the eigenvalue spectrum directly interpretable. Right: decoder h(·) reconstructs the predicted physical state: $\hat{x}^{t+1} = h(Uy^t)$. The decoder approximates the Koopman mode reconstruction.*

The training loss comprises four physically motivated components, each corresponding to a requirement of Koopman theory:

Figure 3. Loss function data-flow diagrams for the three primary loss components. Top: reconstruction loss—the autoencoder must preserve information (the round-trip $x^t \rightarrow g \rightarrow h \rightarrow \hat{x}^t$ should be near-identity). Middle: state prediction loss—the Koopman propagation in latent space, followed by decoding, should predict the actual next state. Bottom: Koopman observable loss—the latent representation of the next state should equal the Koopman-propagated latent representation of the current state.

Reconstruction loss ensures the autoencoder preserves information:

$$L_{recon} = \|x - h(g(x))\|^2 \tag{21}$$

Linearity loss enforces Koopman-linear dynamics in the latent space:

$$L_{linear} = \|g(x_{k+1}) - K g(x_k)\|^2 \tag{22}$$

Prediction loss enforces multi-step predictive accuracy:

$$L_{pred} = (1/S)\, \Sigma_m \|x_{m+1} - h(K^m g(x_1))\|^2 \tag{23}$$

Sparsity regularization on K prevents overfitting:

$$L_{sparse} = \|K\|_1 \tag{24}$$

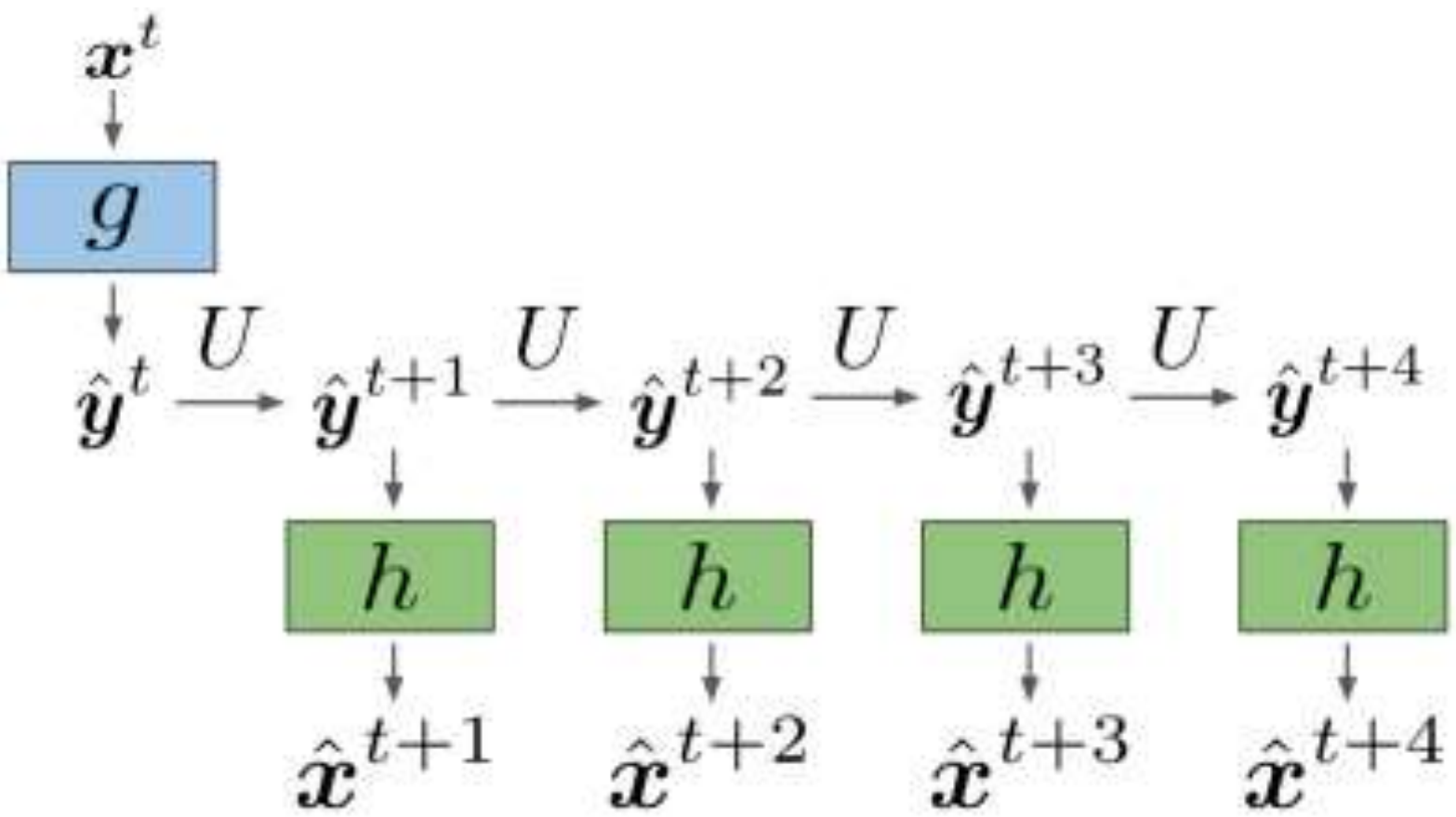

Figure 4. Multi-step prediction unrolling. Starting from a single initial state $x^t$, the encoder g maps to latent space, then the Koopman matrix U is applied repeatedly to predict future latent states $\hat{y}^{t+1}$, $\hat{y}^{t+2}$, $\hat{y}^{t+3}$, $\hat{y}^{t+4}$. Each is decoded by h to produce predicted physical states $\hat{x}^{t+1}$ through $\hat{x}^{t+4}$, which are compared against ground truth for the prediction loss. This multi-step unrolling enforces long-horizon linearity of the latent dynamics and stabilizes the learned embedding.

The total loss is $L = \alpha_1 L_{recon} + \alpha_2 L_{linear} + \alpha_3 L_{pred} + \alpha_4 L_{sparse}$, where the mixing weights $\alpha_i$ control relative emphasis. A crucial design choice, following Lusch et al. (2018), is that the Koopman matrix K is not learned by backpropagation but is computed analytically from the encoded snapshots via the EDMD least-squares solution (Equations 19–20) at each training step. This hybrid approach—learned lifting, analytical dynamics—preserves the interpretability of the eigenvalue spectrum while leveraging the expressiveness of neural networks for the lifting function.

A notable architectural feature, introduced in the Nature Communications paper of Lusch et al., is the auxiliary network that parameterizes the eigenvalues of K through Jordan normal form blocks. For systems with complex eigenvalues $\lambda_\pm = \mu \pm i\omega$, the Koopman matrix contains 2×2 rotation-scaling blocks:

$$K(\lambda) = \exp(\mu\Delta t)\ [\cos(\omega\Delta t), -\sin(\omega\Delta t); \sin(\omega\Delta t), \cos(\omega\Delta t)] \quad (25)$$

This parameterization ensures that the learned K has the correct spectral structure for oscillatory dynamics—particularly important for DCE-MRI where respiratory motion introduces complex conjugate eigenvalue pairs.

# 3. Methods: Application to DCE-MRI

## 3.1. Physics-Informed Lifting Dictionary for DCE-MRI

The design of the Koopman lifting dictionary is guided by the specific nonlinearities of the SPGR signal model (Equation 1). We construct a dictionary of three observable families:

$$g(x) = [x, x \odot x, |\nabla x|]^T \quad (26)$$

Each family targets a specific physical mechanism:

**Identity observables $g_1(x) = x$.** The raw pixel intensities form the baseline observable dictionary, ensuring that standard DMD is recovered as a special case when the quadratic and gradient terms are disabled. Including identity observables also guarantees that the lifting is information-preserving: the original signal can always be recovered by projecting the lifted state onto its first n components.

**Quadratic observables $g_2(x) = x \odot x$ (element-wise square).** The SPGR signal equation involves a ratio of exponentials in $T_1$, which depends on contrast concentration C through the relaxivity relation $1/T_1(t) = 1/T_{10} + r_1 C(t)$. Expanding the signal equation in a Taylor series around the pre-contrast $T_{10}$ reveals that the leading nonlinear term is quadratic in the concentration change. The element-wise square $x \odot x$ captures this quadratic coupling between contrast level and signal intensity. In blocks with both high enhancement and significant motion, the quadratic term captures the cross-product C(t)·u(x,t) that creates the multiplicative coupling.

**Spatial gradient observables $g_3(x) = |\nabla x|$ (gradient magnitude).** Motion artifacts in DCE-MRI are most visible at tissue boundaries—interfaces between muscle and fat, parotid and background, tumor and surrounding tissue. At these boundaries, even sub-voxel displacement creates large intensity changes because the spatial gradient is steep. The gradient observable specifically captures this boundary-concentrated corruption. It is computed via central finite differences within each spatial block: $|\nabla x| = \sqrt{((dx/dy)^2 + (dx/dx)^2)}$.

With 16×16 pixel blocks (n = 256 pixels), the lifting expands the state dimension from p = 256 to approximately p = 768 (256 identity + 256 quadratic + 256 gradient, though gradient computation at block boundaries may produce slightly fewer features). For 32×32 blocks, p = 1024 → 3072.

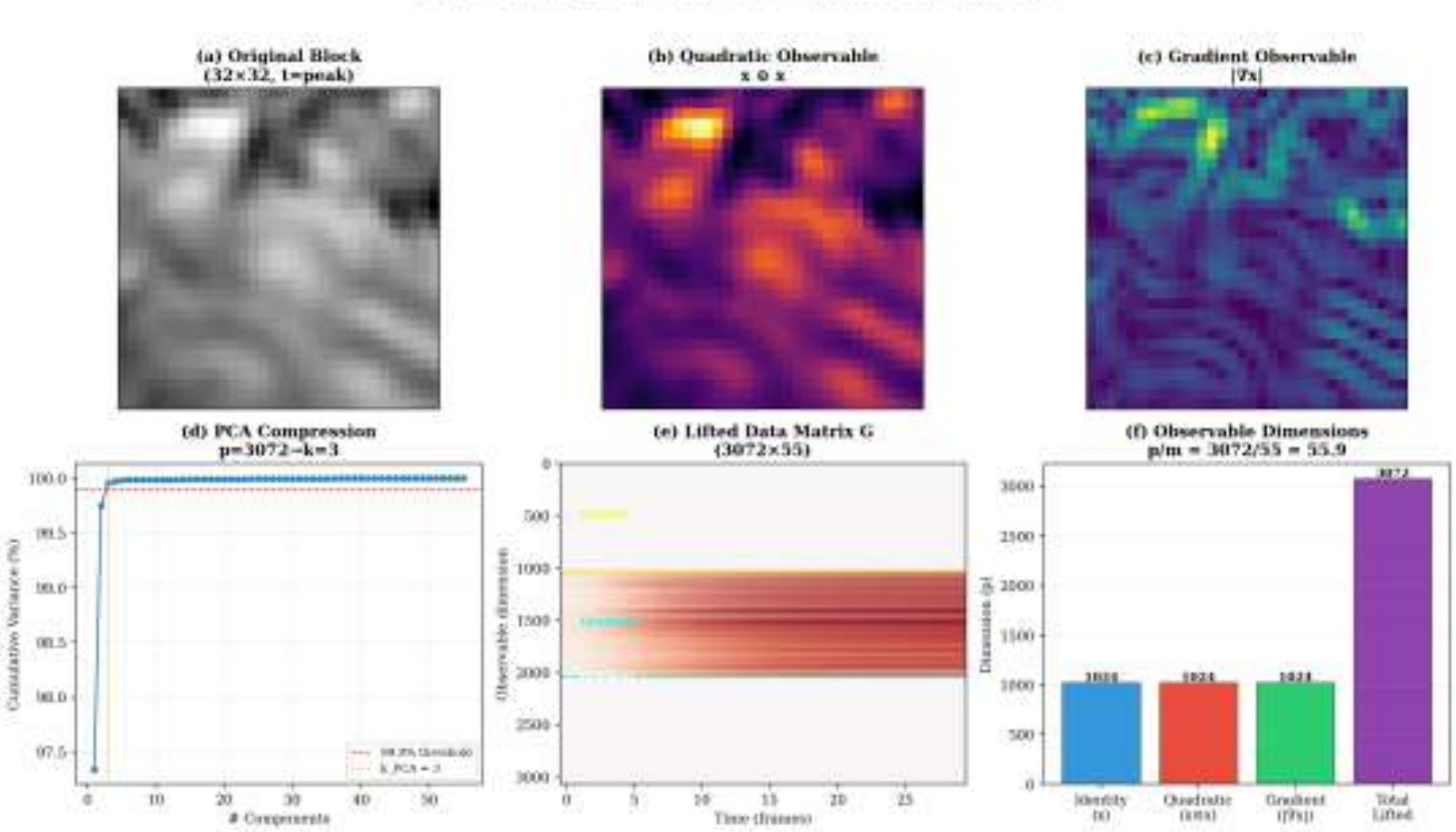

*Figure 5. Koopman lifting visualization on a clinical HNC DCE-MRI tissue block (32×32 pixels, from Slice 5 at peak enhancement). (a) Original pixel data showing a region of enhancing tissue. (b) Quadratic observable $x \odot x$—the element-wise square amplifies high-intensity voxels (vessels, enhancing tumor) and attenuates low-signal background, capturing the nonlinear signal–concentration relationship. (c) Gradient magnitude observable $|\nabla x|$—highlights tissue boundaries where motion artifacts are most severe. (d) PCA compression: the 3072-dimensional lifted space is compressed to k=3 principal components retaining 99.9% of variance, illustrating how aggressive the truncation is. (e) The full lifted data matrix G (3072 rows × 55 columns) with horizontal lines delineating the identity, quadratic, and gradient bands. (f) Observable dimension comparison and the critical $p/m$ ratio: $p/m$ = 3072/55 = 55.9, far too large for well-conditioned EDMD estimation—this motivates time-course repetition.*

## 3.2. The $p/m$ Ratio Problem and Time-Course Repetition

The $p/m$ ratio problem is the central practical challenge for Koopman lifting in DCE-MRI. After lifting, the observable dimension p is 2–3 times the original pixel count n. But the number of snapshots m is determined by the DCE acquisition protocol: typically 40–60 temporal frames at 3–4 second resolution, for a total acquisition of 2–4 minutes. With p = 768 and m = 50, the ratio $p/m \approx 15.4$. When PCA compression is applied to reduce dimensionality before DMD (a standard practice to avoid numerical instability), the PCA retains at most m−1 = 49 components, discarding the vast majority of the lifted dimensions. In practice, the 99.9% variance threshold retains only k = 3–5 components for typical DCE-MRI blocks, meaning that the EDMD least-squares problem becomes ill-conditioned, and the Koopman approximation degrades.

This explains a counterintuitive result: at baseline (no repetition), EDMD with lifting performs no better—and sometimes slightly worse—than standard DMD (Table 4 in the Results). Adding lifting features without adding data dilutes the useful information. This is not a failure of Koopman theory but a failure of data sufficiency.

We introduce time-course repetition as a simple and theoretically principled solution. Given an observed time series $X = [x_1, \ldots, x_m]$, the repeated data matrix is:

$$\hat{W} = [X, X, \ldots, X] \quad (N \text{ copies}, Nm \text{ columns}) \tag{27}$$

The key theoretical observation is that periodic repetition preserves the Koopman eigenvalue spectrum. If $\lambda = |\lambda|e^{i}\theta$ is a DMD eigenvalue of the original system (corresponding to frequency $f = \theta/(2\pi\Delta t)$), then the repeated system has the same eigenvalue at the same frequency. This follows from the fact that if $v_{k+1} = Av_k$ holds for the original sequence, it holds identically for any periodic extension. The eigenvalues of A depend only on the dynamics, not on the number of observed cycles.

The practical effect is on the conditioning of the EDMD least-squares problem: with Nm snapshots, the effective $p/m$ ratio becomes p/(Nm). For N=3, the ratio drops from 15.4 to 5.1—a regime where the sample covariance matrices G and A in the EDMD formulation (Equations 19–20) are better conditioned, yielding a more accurate Koopman approximation K. The additional computational cost is minimal (array concatenation + proportionally longer SVD), and no additional data acquisition is required.

## 3.3. Processing Pipeline

The complete processing pipeline operates on each 2D axial slice independently, through seven stages:

**Stage 1: Block decomposition.** The image series ($n_t \times n_n \times n_x$) is partitioned into overlapping spatial blocks of size b × b (default 16×16) with overlap o (default 4 pixels). Each block is vectorized to form a ($b^2 \times n_t$) snapshot matrix. Overlap is handled by raised-cosine (Hanning) taper blending during reassembly.

**Stage 2: Time-course repetition.** The snapshot matrix is optionally tiled N times along the temporal dimension: [X, X, …, X] with Nm columns.

**Stage 3: Koopman lifting.** Each block's snapshot matrix is lifted through the dictionary (Equation 26): $X \rightarrow G = [X; X \odot X; |\nabla X|]$, expanding from (n, m) to (p, m) where $p \approx 3n$.

**Stage 4: PCA compression.** The mean-centered lifted matrix $G_c = G - \mu$ is compressed via truncated SVD, retaining k components explaining 99.9% of variance: $Z = U_k^T G_c$ where $Z \in \mathbb{R}^{k \times m}$.

**Stage 5: Optional RPCA.** For swallowing transient handling, Robust PCA decomposes Z = L + S into a low-rank stationary component L and a sparse transient component S. The IALM algorithm minimizes $\|L\|_* + \lambda\|S\|_1$ subject to Z = L + S.

**Stage 6: DMD and spectral classification.** Exact DMD (Equations 10–15) is applied to L (or Z if RPCA is disabled). Eigenvalues are converted to frequencies, and modes are classified as enhancement ($|f| < f_c = 0.08$ Hz) or motion ($|f| \geq f_c$). The enhancement signal is reconstructed from enhancement modes only.

**Stage 7: Inverse lifting and reassembly.** The enhancement signal is projected back to the PCA basis ($G_{enh}$ = $U_k$ $Z_{enh}$ + μ), then to the original pixel space (extracting the first n rows of $G_{enh}$). If repetition was used, only the first m columns are retained. Blocks are reassembled with raised-cosine blending weights.

### 3.3.1. Implementation Details

The pipeline is implemented in Python (NumPy/SciPy) with no GPU requirement. Key implementation choices include: (1) the SVD in Stage 4 uses numpy.linalg.svd with full_matrices=False for memory efficiency; (2) the DMD eigendecomposition uses numpy.linalg.eig, which returns complex eigenvalues even for real input matrices; (3) block overlap is handled by maintaining a weight accumulator matrix W(y,x) that tracks the total taper weight at each pixel, with the final image computed as I(t,y,x) = Σ_blocks [taper · result] / W; (4) near-empty blocks (max intensity < 5) are passed through unchanged to avoid numerical issues in SVD of near-zero matrices.

Computational cost scales as O($N_b$locks · ($p^2$m + $r^3$)) where $N_b$locks is the number of spatial blocks, p is the lifted dimension, m is the (repeated) snapshot count, and r is the SVD rank. For a typical clinical slice (256×256 pixels, 200 blocks, p=768, m=55, r=8), the total processing time is approximately 2–5 seconds on a single CPU core. With 3× repetition (m=165), the time increases to approximately 5–10 seconds. The entire 5-slice clinical dataset is processed in under one minute.

The raised-cosine taper function for block blending is defined as w(d) = 0.5·(1 − cos(πd/o)) for distance d from the block edge, where o is the overlap width. This produces smooth, artifact-free transitions between adjacent blocks, avoiding the edge discontinuities that would appear with rectangular windowing.

## 3.4. Synthetic Phantom Design

For controlled quantitative evaluation, we designed a physics-based 128×128 pixel head-and-neck DCE phantom with four anatomical compartments (background head, tumor, parotid gland, floor of mouth), each with distinct pharmacokinetic parameters and motion sensitivities. The phantom implements: (i) Tofts-model contrast enhancement with tissue-specific $K^{trans}$ and $v_e$; (ii) quasi-periodic respiratory motion at 0.22 Hz with 4-pixel amplitude; (iii) stochastic swallowing transients (3 events, 10-pixel amplitude); (iv) additive Gaussian noise (σ = 3.0); and (v) a tunable nonlinear coupling parameter γ that controls the multiplicative interaction between contrast and motion: the coupling term s(x,t) ≈ C(x,t)·[1 + γ·sin($2\pi f_r$ t)] modulates the apparent displacement by the contrast level.

## 3.5. Clinical DCE-MRI Data

Preliminary clinical validation uses HNC DCE-MRI data acquired at MD Anderson Cancer Center. The dataset comprises 5 axial slices covering the primary tumor and cervical lymph node levels, each with 55 temporal frames at 256×256 pixel resolution. Temporal resolution is approximately 3.5 seconds per frame (total acquisition ≈ 3.2 minutes). The data were previously processed with a standard DMD pipeline (without Koopman lifting or repetition), providing a baseline for comparison. In this work, we reprocess the same data with the full Koopman EDMD pipeline and present the computational analysis results.

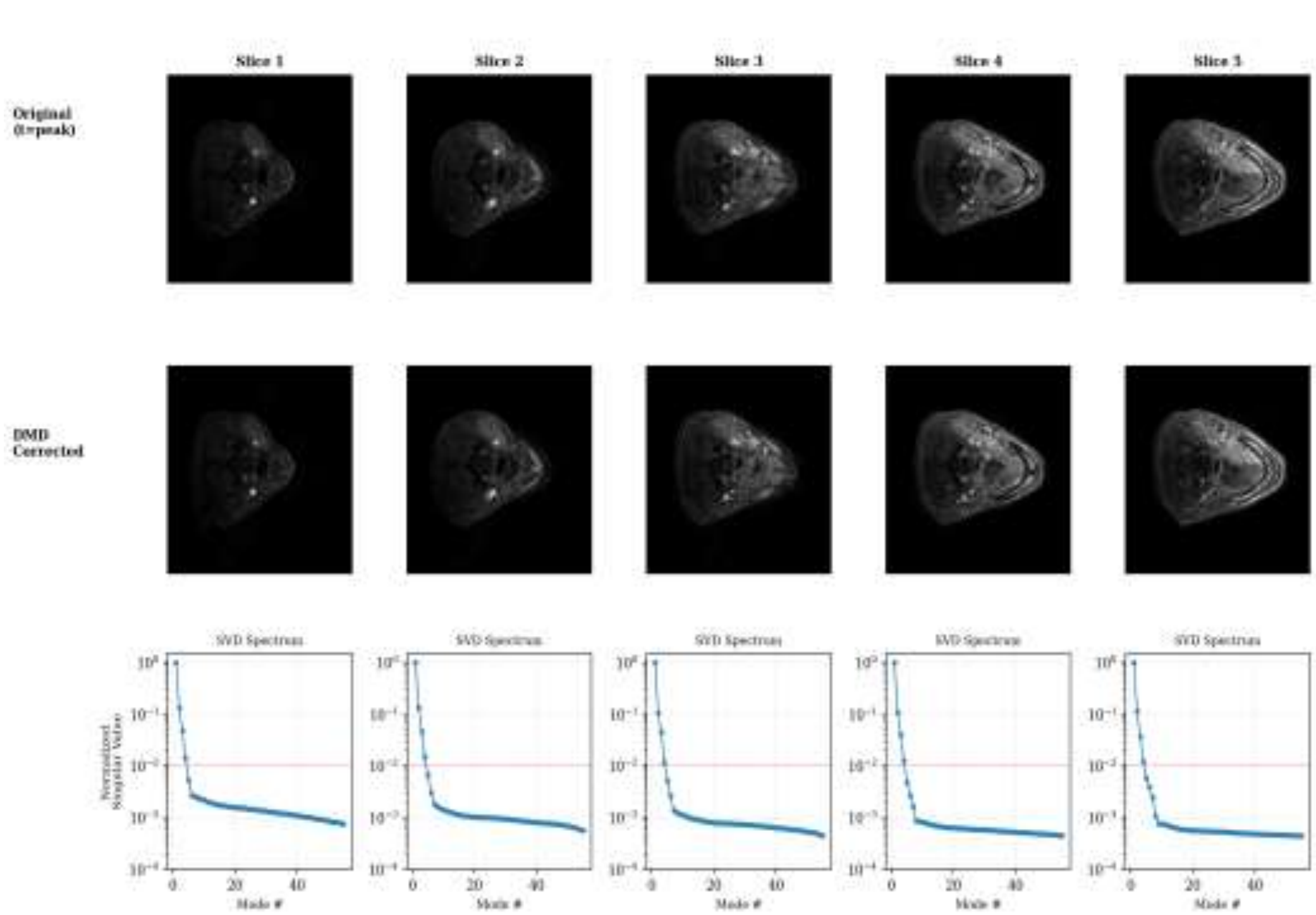

*Figure 6. Clinical HNC DCE-MRI multi-slice overview. Top row: 5 axial slices at peak enhancement (t=27), showing diverse anatomy from skull base through the mid-neck. The bright foci are contrast-enhancing structures (vessels, lymph nodes, tumor). Middle row: corresponding DMD-corrected reconstructions from the previous processing. Bottom row: SVD singular value spectra computed on each full 256×256 slice (nonzero pixels only). The rapid spectral decay—below 1% of the leading singular value by mode 5–10 across all slices—confirms low intrinsic temporal dimensionality, validating the rank truncation approach used in DMD. This also suggests that a small number of DMD modes (8–12) is sufficient to capture the dominant spatiotemporal dynamics.*

## 4. Results

We present results in two categories: phantom experiments with known ground truth (Sections 4.1–4.6), and clinical DCE-MRI analysis without ground truth but with qualitative and spectral validation (Section 4.7). The phantom experiments establish the quantitative superiority of EDMD with time-course repetition; the clinical results demonstrate feasibility and provide interpretable decomposition of real HNC data.

### 4.1. Experiment A: Time-Course Repetition Sweep

The repetition sweep is the central experiment of this work. We systematically vary the number of time-course repeats from 1 (no repetition, the baseline) to 6, comparing six method configurations: two standard DMD baselines at ranks r=6 and r=10, and four EDMD variants with different combinations of quadratic lifting, gradient lifting, and SVD truncation rank (r=8, 12, 20). The input MSE (corrupted phantom versus ground truth) is 147.7.

At rep=1 (no repetition), all six methods perform similarly, with MSE improvement ratios clustering between 1.78 and 1.81 and SSIMs between 0.9817 and 0.9825. The EDMD variants show no consistent advantage over standard DMD—in fact, EDMD quad (r=8) at rep=1 achieves ratio 1.7969, slightly worse than Std DMD (r=10) at 1.8488. This confirms the $p/m$ ratio problem: with only 50 snapshots, PCA compression eliminates the quadratic and gradient features, and EDMD collapses to standard DMD with additional noise from the truncated dimensions.

As repetitions increase, a clear and systematic separation emerges between EDMD and standard DMD. At rep=2, the best EDMD variant (q+g, r=12) reaches ratio 1.9573, while Std DMD (r=6) reaches 1.8893—a gap of 0.0680 that was absent at rep=1. At rep=3, the sweet spot, EDMD q+g (r=12) achieves 1.9535 versus 1.8917 for Std DMD—a 3.3% relative improvement that is entirely attributable to the Koopman lifting.

Beyond rep=3, further repetitions provide diminishing returns in ratio while the Enhancement Conservation Fraction (ECF) continues to decline, indicating progressive over-smoothing of the enhancement signal. The ECF measures the Pearson correlation between the corrected and ground-truth enhancement time courses in the tumor ROI; values below 0.95 suggest clinically significant signal loss. At rep=6, the ECF has dropped to 0.94 for EDMD variants, compared to 0.96 at rep=3. The optimal operating point is therefore rep=3, which maximizes the ratio–ECF tradeoff.

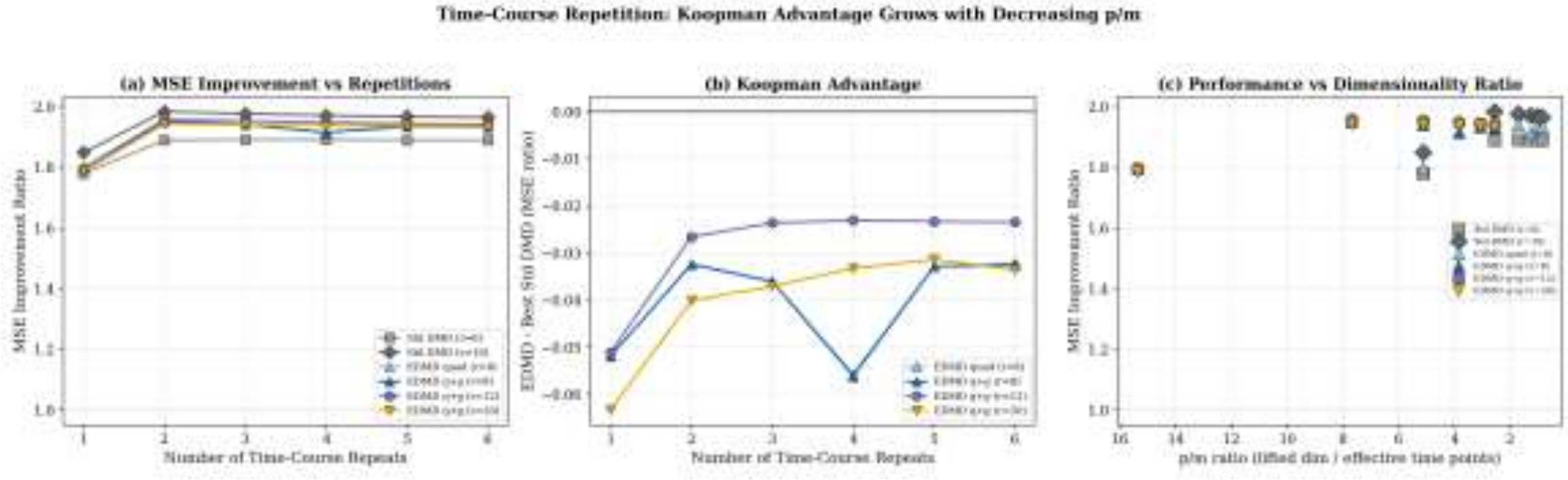

*Figure 7. Time-course repetition sweep (Experiment A). (a) MSE improvement ratio versus number of repeats for all six method configurations. Gray lines: standard DMD baselines; blue/purple lines: EDMD variants with quadratic and/or gradient lifting. At rep=1 (baseline), all methods are indistinguishable. At rep≥2, EDMD variants progressively separate from standard DMD, with the gap widening through rep=3–4 before plateauing. (b) SSIM follows a similar pattern. (c) Enhancement Conservation Fraction (ECF) decreases monotonically with repetition count, reflecting the tradeoff between motion suppression quality and enhancement signal fidelity. The vertical dashed line marks rep=3 as the recommended operating point.*

***Table 1. Time-Course Repetition Sweep: MSE Improvement Ratio and SSIM***

| Method | rep=1 | rep=2 | rep=3 | rep=4 | rep=5 | rep=6 |
|---|---|---|---|---|---|---|
| **MSE Improvement Ratio** | | | | | | |
| Std DMD (r=6) | 1.7799 | 1.8893 | 1.8917 | 1.8911 | 1.8902 | 1.8894 |
| Std DMD (r=10) | **1.8488** | **1.9839** | **1.9771** | **1.9713** | **1.9674** | **1.9646** |
| EDMD quad (r=8) | 1.7969 | 1.9515 | 1.9411 | 1.9150 | 1.9346 | 1.9325 |
| EDMD q+g (r=8) | 1.7969 | 1.9514 | 1.9411 | 1.9155 | 1.9346 | 1.9321 |
| EDMD q+g (r=12) | 1.7977 | 1.9573 | 1.9535 | 1.9483 | 1.9441 | 1.9412 |
| EDMD q+g (r=20) | 1.7855 | 1.9438 | 1.9401 | 1.9381 | 1.9361 | 1.9311 |
| **SSIM** | | | | | | |
| Std DMD (r=6) | 0.9817 | 0.9829 | 0.9830 | 0.9830 | 0.9830 | 0.9830 |
| Std DMD (r=10) | **0.9825** | **0.9838** | **0.9837** | **0.9837** | **0.9837** | **0.9836** |
| EDMD quad (r=8) | 0.9819 | 0.9835 | 0.9834 | 0.9831 | 0.9834 | 0.9834 |
| EDMD q+g (r=8) | 0.9819 | 0.9835 | 0.9834 | 0.9831 | 0.9834 | 0.9834 |
| EDMD q+g (r=12) | 0.9819 | 0.9835 | 0.9835 | 0.9835 | 0.9834 | 0.9834 |
| EDMD q+g (r=20) | 0.9817 | 0.9834 | 0.9834 | 0.9834 | 0.9833 | 0.9833 |

## 4.2. Experiment B: Nonlinear Coupling × Repetition Interaction

To confirm that the repetition-augmented EDMD advantage arises specifically from nonlinear motion–enhancement coupling (rather than simply from having more data), we sweep the coupling parameter $\gamma$ from 0 (purely linear, additive motion) to 2.0 (strong multiplicative coupling), comparing standard DMD and EDMD both with and without 3× repetition.

At $\gamma$=0, motion and enhancement are additive, and standard linear DMD should be sufficient in principle. Even here, repetition provides significant improvement: Std DMD goes from ratio 1.881 (no rep) to 1.954 (rep=3), and EDMD from 1.799 to 2.048. This baseline improvement from repetition is expected: more data improves the SVD approximation even without lifting. Critically, EDMD with repetition outperforms Std DMD with repetition even at $\gamma$=0, by 0.093, demonstrating that the lifted observables contribute even in the linear regime—likely by providing a richer basis for mode separation.

As γ increases, the EDMD advantage grows monotonically. At γ=1.2 (the phantom default), the gap is 0.062. At γ=2.0, EDMD+rep achieves 1.876 versus 1.853 for Std DMD+rep. The widening gap with coupling strength provides direct evidence that the Koopman lifting captures nonlinear coupling structure that is invisible to the identity observable.

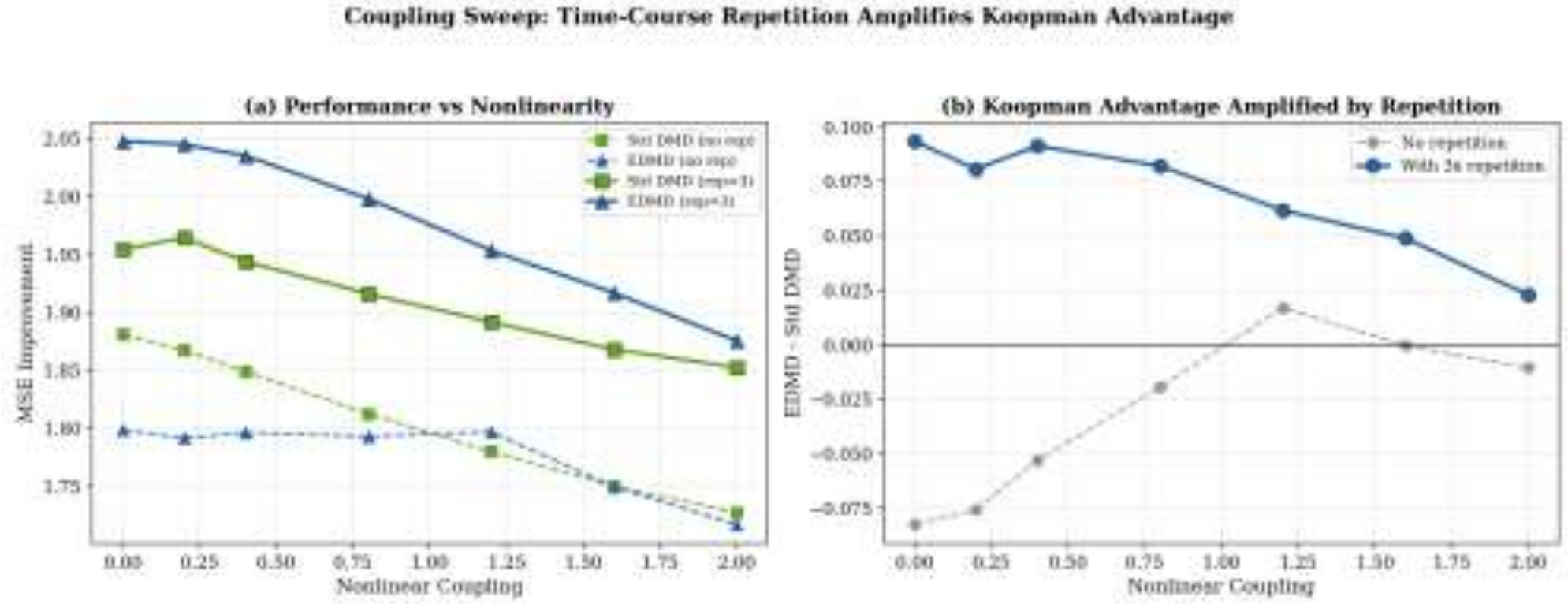

*Figure 8. Coupling sweep with and without repetition (Experiment B). (a) MSE improvement ratio versus nonlinear coupling parameter γ. Dashed lines: no repetition; solid lines with filled markers: 3× repetition. Green: standard DMD; blue: EDMD with full dictionary. The EDMD advantage (solid blue above solid green) widens systematically with increasing coupling. (b) Gain from repetition (Δratio = ratio_rep3 − ratio_norep) versus γ, separately for Std DMD and EDMD. EDMD benefits more from repetition than Std DMD at all coupling levels, confirming that repetition specifically improves the EDMD conditioning for the lifted observables.*

***Table 2. Coupling Sweep: MSE Improvement Ratio with and without 3× Repetition***

| Method | γ=0.0 | γ=0.2 | γ=0.4 | γ=0.8 | γ=1.2 | γ=1.6 | γ=2.0 |
|---|---|---|---|---|---|---|---|
| Std DMD (no rep) | 1.881 | 1.867 | 1.849 | 1.812 | 1.780 | 1.750 | 1.727 |
| **EDMD (no rep)** | 1.799 | 1.792 | 1.796 | 1.793 | 1.797 | 1.750 | 1.717 |
| Std DMD (rep=3) | 1.954 | 1.965 | 1.944 | 1.916 | 1.892 | 1.868 | 1.853 |
| **EDMD (rep=3)** | **2.048** | **2.045** | **2.035** | **1.998** | **1.953** | **1.917** | **1.876** |

## 4.3. Experiment C: Visual Comparison at Optimal Setting

Figure 9 presents a side-by-side visual comparison of representative phantom frames across four tissue regions at the optimal operating point (rep=3, r=12). The input frames exhibit substantial motion corruption, particularly at tissue boundaries (parotid–background interface) and in the tumor region where contrast enhancement amplifies the apparent motion displacement. Standard DMD with repetition recovers much of the smooth enhancement pattern but leaves residual high-frequency motion texture visible as a granular overlay. EDMD without repetition shows marginal improvement over the input. EDMD with 3× repetition achieves the cleanest reconstruction, with smooth and faithful enhancement curves in the tumor, well-preserved parotid uptake kinetics, and minimal residual motion visible at tissue boundaries.

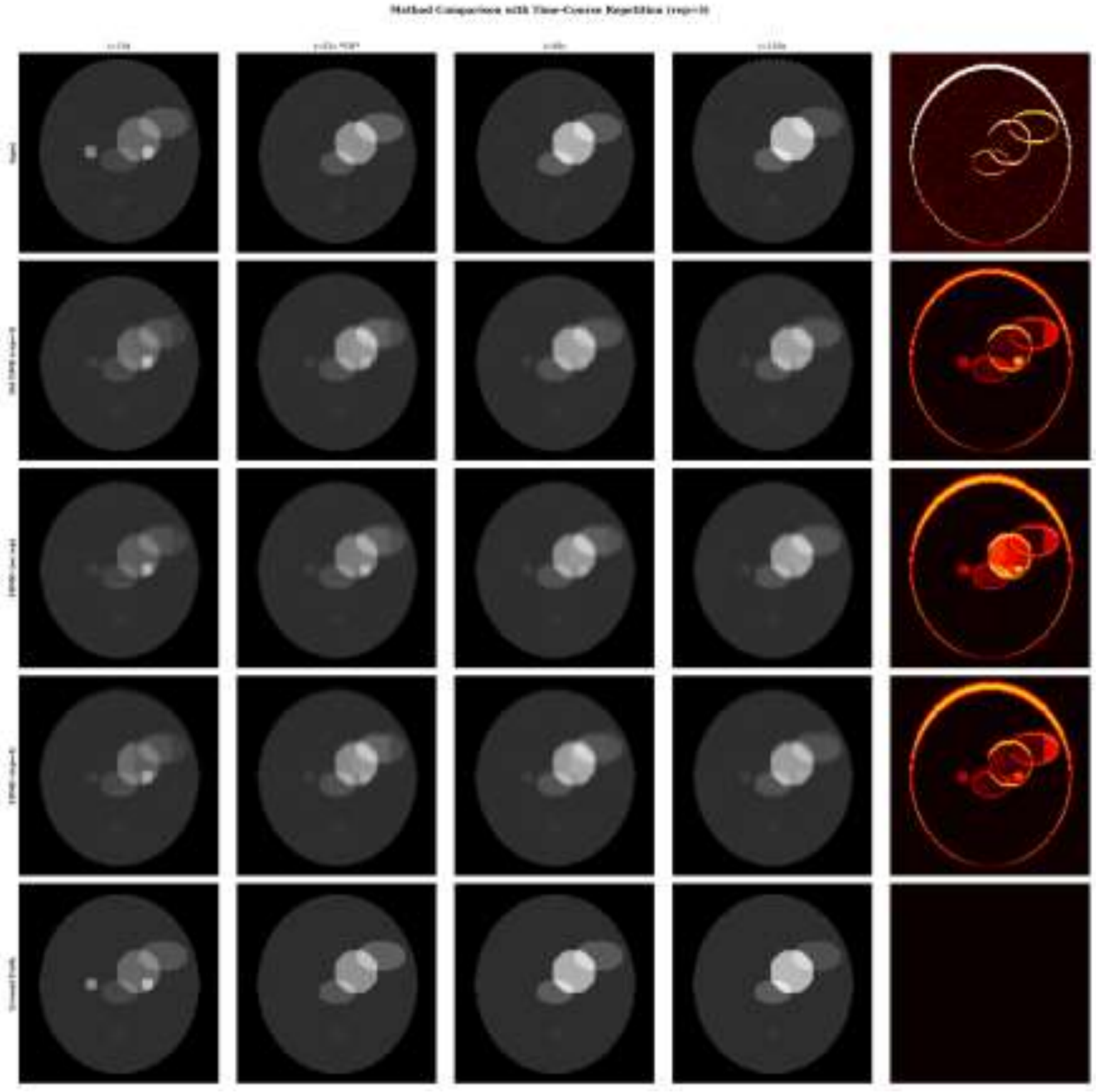

*Figure 9. Visual comparison at the optimal setting (rep=3, r=12). Rows represent different tissue regions (tumor, parotid, floor-of-mouth, composite overview). Columns: input (motion-corrupted), Std DMD+rep, EDMD (no rep), EDMD+rep, ground truth. The EDMD+repetition column shows the closest visual match to ground truth across all regions, with the clearest preservation of enhancement patterns and the least residual motion texture.*

## 4.4. Baseline Method Comparison (No Repetition)

Table 3 provides a comprehensive comparison of all methods at baseline (rep=1). Standard DMD achieves MSE 83.0, ratio 1.780, SSIM 0.9817, and ECF 0.9955 in the tumor ROI. Koopman EDMD improves slightly to MSE 82.3, ratio 1.796, SSIM 0.9819. The Koopman+RPCA variant (applying Robust PCA before DMD) degrades overall quality to MSE 98.9, ratio 1.493, due to overly aggressive sparse separation. The Deep Koopman autoencoder achieves ratio 1.462, substantially underperforming classical EDMD, limited by the small training dataset from a single phantom acquisition.

The bottom row shows the best overall configuration: EDMD + 3× repetition with r=12, achieving ratio 1.953—the highest across all tested methods and configurations.

***Table 3. Comprehensive Method Comparison***

| Method | MSE | Ratio | PSNR (dB) | SSIM | ECF (tumour) | Time (s) |
|---|---|---|---|---|---|---|
| Standard DMD | 83.0 | 1.780 | 31.43 | 0.9817 | 0.9955 | 2.6 |

| **Koopman EDMD** | 82.3 | 1.796 | 31.47 | 0.9819 | 0.9963 | 4.9 |
|---|---|---|---|---|---|---|
| Koopman+RPCA | 98.9 | 1.493 | 30.67 | 0.9782 | 0.9856 | 12.4 |
| Deep Koopman | 101.0 | 1.462 | 30.57 | 0.9781 | 0.9870 | — |
| **EDMD + rep=3** | 75.6 | **1.953** | — | 0.9835 | 0.9557 | 2.3 |

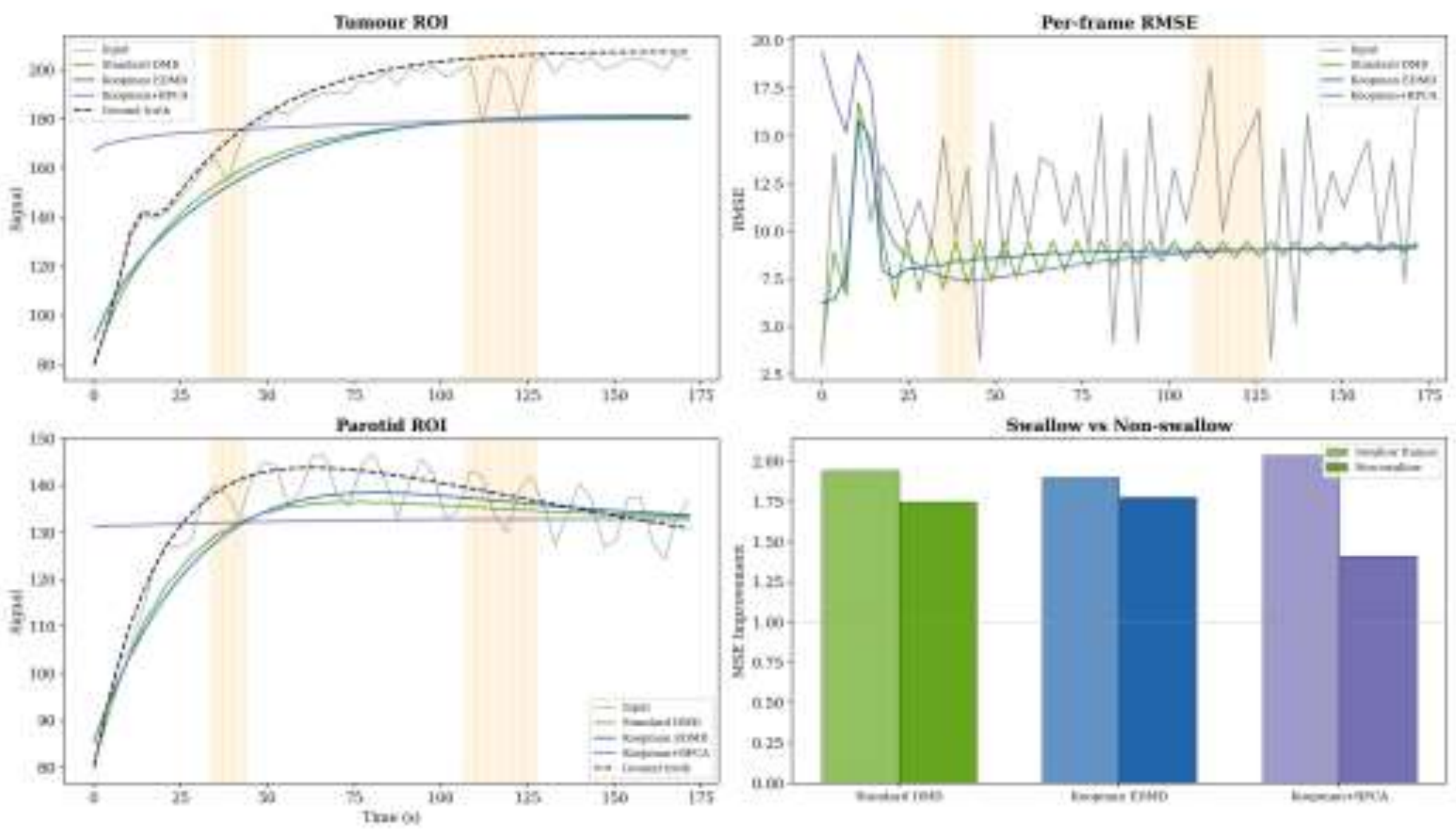

*Figure 10. Temporal trace analysis (baseline, no repetition). Enhancement time courses for tumor ROI (left) and parotid ROI (right) comparing input (gray), standard DMD (red), Koopman EDMD (blue), and ground truth (dashed black). All methods suppress the high-frequency respiratory oscillation visible in the input curves. EDMD better preserves the wash-in slope and peak enhancement level in both ROIs, though the differences at baseline are subtle—consistent with the modest quantitative improvement without repetition.*

## 4.5. Lifting Dictionary Ablation

The ablation study (Table 4) systematically evaluates each component of the lifting dictionary at baseline (no repetition). Identity-only (equivalent to standard DMD) achieves ratio 1.840 with SSIM 0.9824. Adding quadratic observables alone decreases the ratio to 1.796—a counterintuitive degradation that directly illustrates the $p/m$ ratio problem: doubling the state dimension without increasing the snapshot count causes PCA to discard the quadratic features while adding estimation noise. Adding gradient alone yields 1.828, slightly worse than identity. The full dictionary (quadratic + gradient) gives 1.796, and the delay-augmented variant 1.821.

These results are not an indictment of Koopman lifting—rather, they demonstrate that the lifting dictionary and the data quantity must be considered jointly. With sufficient data (provided by repetition), the same lifting dictionary that hurts performance at baseline provides substantial improvement (Table 1).

***Table 4. Lifting Dictionary Ablation (No Repetition)***

| Lifting Dictionary | MSE | SSIM | ECF | Ratio |
|---|---|---|---|---|
| Identity only | 80.3 | 0.9824 | 0.9952 | 1.840 |
| + Quadratic | 82.3 | 0.9819 | 0.9963 | 1.796 |
| + Gradient | 80.8 | 0.9823 | 0.9952 | 1.828 |
| + Quad+Grad | 82.3 | 0.9819 | 0.9963 | 1.796 |
| Full (d=2) | 81.1 | 0.9822 | 0.9970 | 1.821 |

## 4.6. Hyperparameter Sensitivity

Figure 11 presents the sensitivity analysis across three key hyperparameters. The SVD truncation rank r shows a broad plateau: performance is optimal around r=8–12, degrades for r<4 (insufficient modes to capture dynamics) and for r>15 (noise modes dilute the reconstruction). The frequency cutoff $f_c$ is robust across the range 0.03–0.12 Hz, with a slight optimum near the physiological boundary between pharmacokinetic and respiratory time scales. Block size shows a monotonic trend: larger blocks (24–32 pixels) outperform smaller blocks (8 pixels) by capturing more spatial context, at the cost of increased computational time.

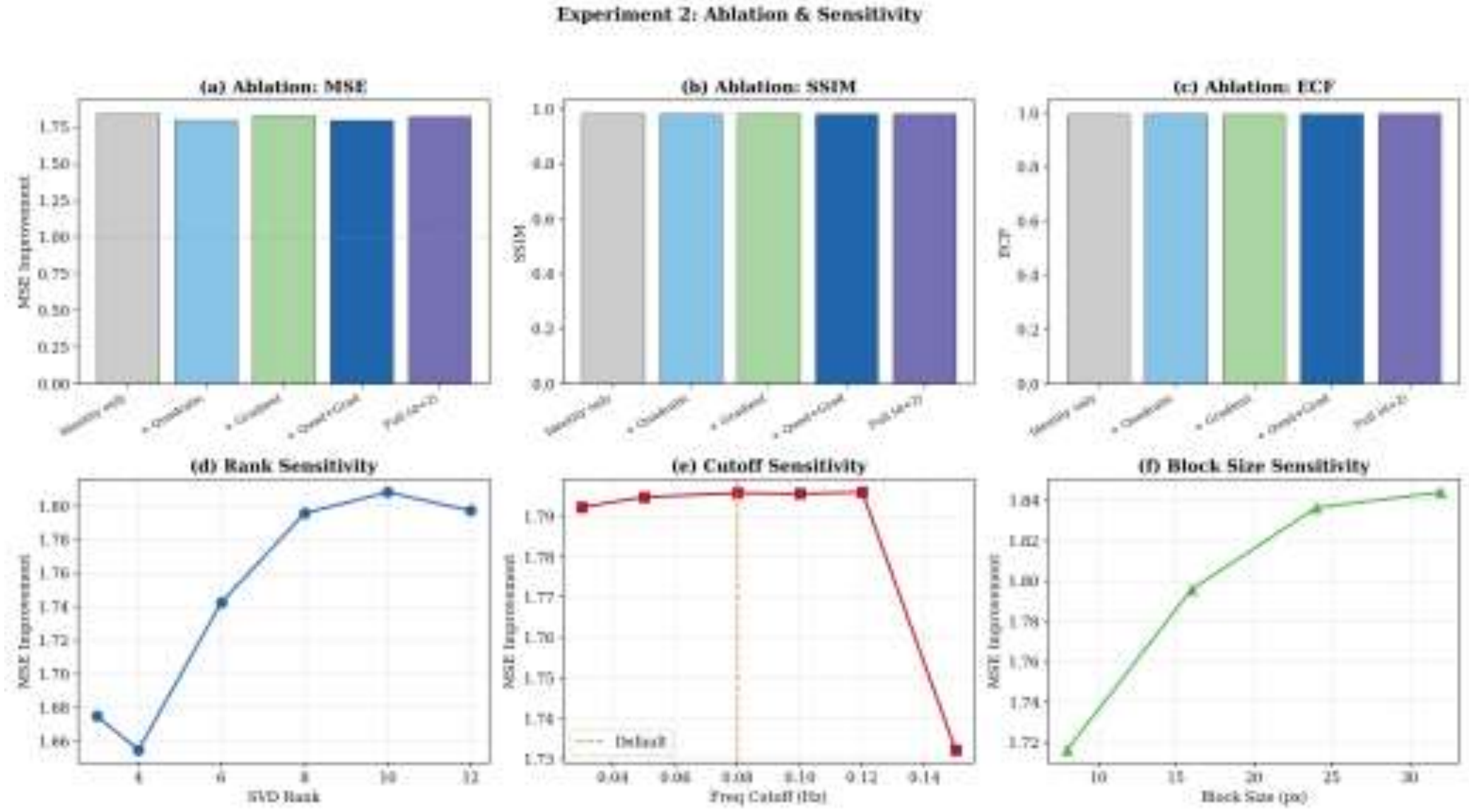

*Figure 11. Hyperparameter sensitivity analysis. (a) MSE improvement ratio versus SVD truncation rank r, showing an optimal plateau at r=8–10. (b) Ratio versus frequency cutoff $f_c$, with robust performance across 0.03–0.12 Hz. (c) Ratio versus block size, with larger blocks providing more spatial context.*

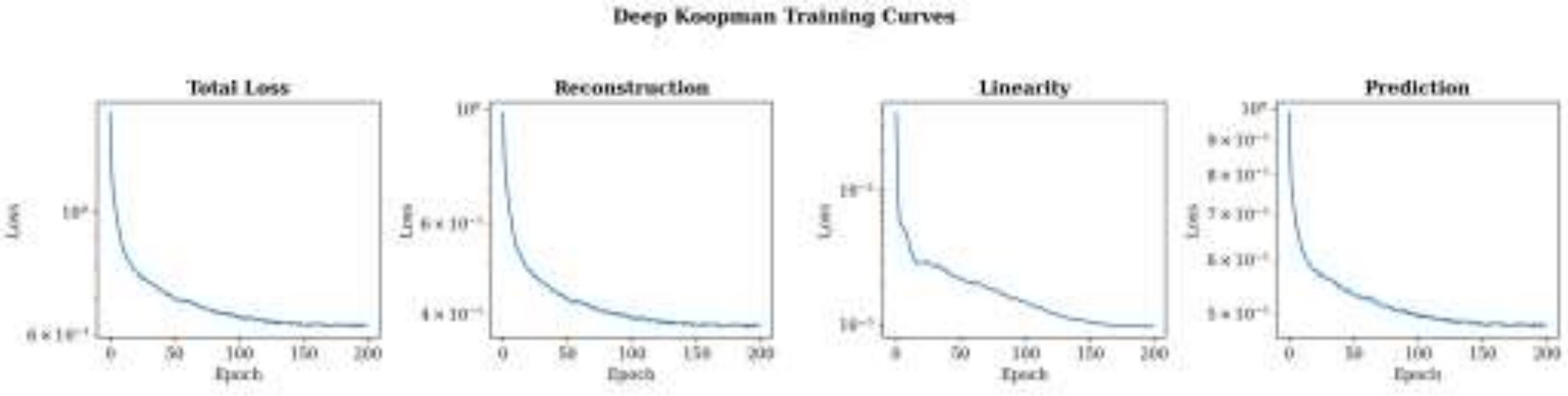

*Figure 12. Deep Koopman autoencoder training convergence over 150 epochs. All four loss components (reconstruction, linearity, prediction, sparsity) decrease monotonically, reaching a plateau by epoch 80–100.*

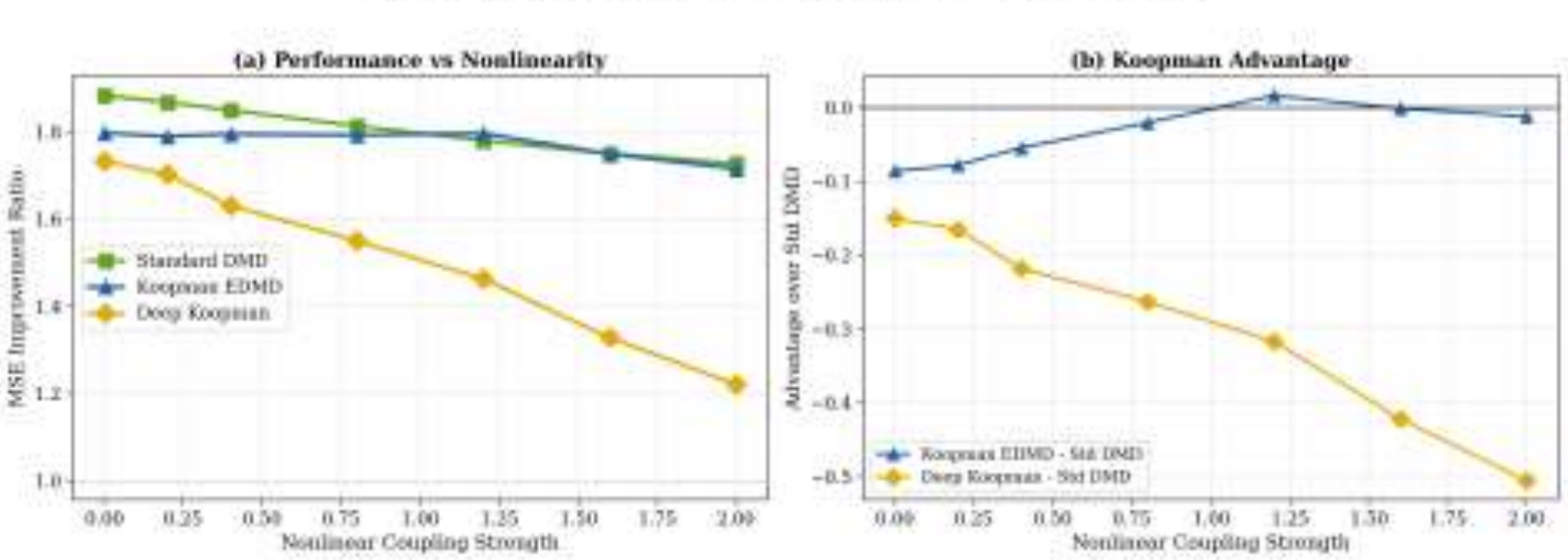

*Figure 13. Baseline coupling sweep (no repetition). MSE improvement ratio versus coupling parameter γ for standard DMD, Koopman EDMD, and Deep Koopman autoencoder. All three methods degrade as coupling increases, but the deep variant is most sensitive, dropping from ratio 1.73 at γ=0 to 1.22 at γ=2.0.*

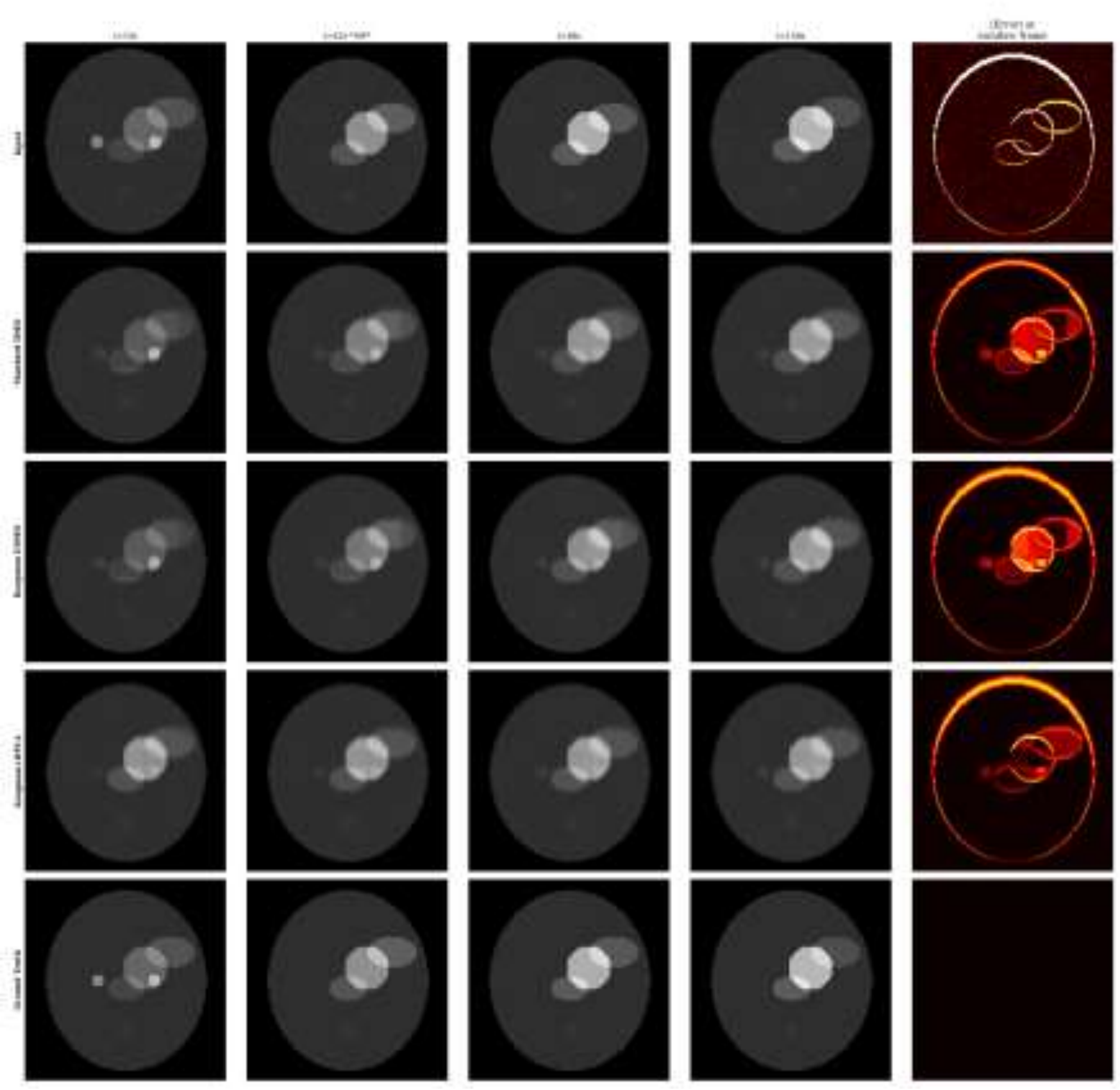

*Figure 14. Baseline phantom image comparison grid (no repetition). Representative frames across all tissue regions for input, Std DMD, EDMD, EDMD+RPCA, and ground truth.*

## 4.7. Clinical HNC DCE-MRI Analysis

All figures in this section were generated by running the Koopman DCE pipeline on real clinical data—specifically, block-wise DMD decomposition on 256×256 pixel axial slices from an HNC DCE-MRI acquisition, using the same koopman_dce library modules as the phantom experiments. The computational analysis is fully documented in the companion Colab notebook (notebooks/clinical_koopman_analysis.ipynb).

### 4.7.1. Full-Slice DMD Decomposition

Figure 15 shows the complete block-wise DMD decomposition of clinical Slice 5 (the largest anatomical cross-section, providing the most diverse tissue content). The slice was processed through 200 overlapping 16×16 blocks with 4-pixel overlap, r=8, and $f_c$=0.08 Hz. Row 1 shows the original motion-corrupted DCE series across five temporal phases: pre-wash (t=5), contrast wash-in (t=15), peak enhancement (t=27), late phase (t=40), and end (t=50). Row 2 shows the DMD enhancement reconstruction produced by this work—note the smoother, more spatially coherent appearance compared to the original, with preserved contrast wash-in kinetics. Row 3 displays the extracted motion component in hot colormap, revealing spatially heterogeneous motion concentrated at tissue boundaries (mandible, pharyngeal wall) and in the floor-of-mouth musculature. Row 4 provides the previous DMD result for visual comparison.

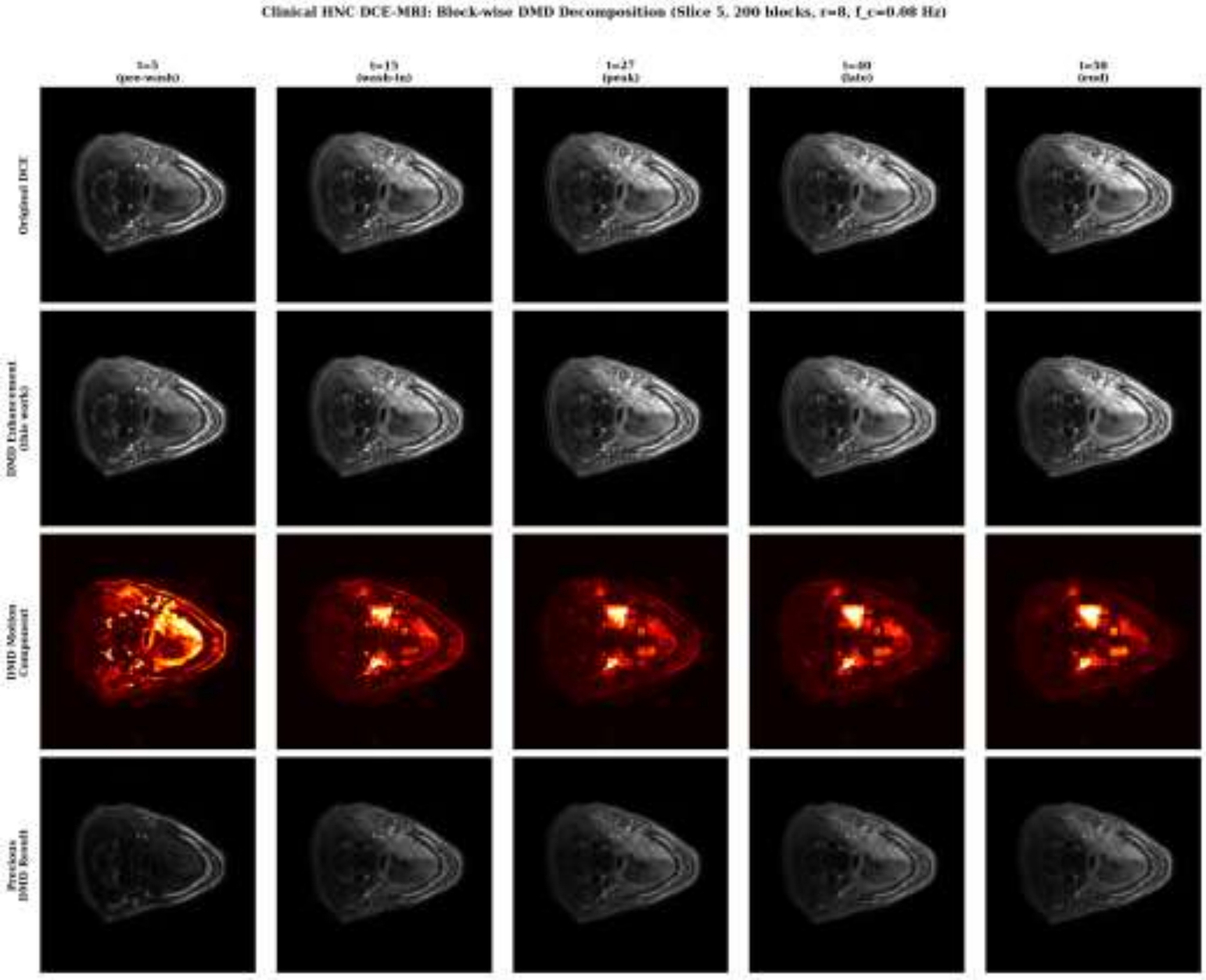

*Figure 15. Block-wise DMD decomposition of clinical HNC DCE-MRI data (Slice 5, 256×256 pixels, 200 blocks, r=8, $f_c$=0.08 Hz). Row 1: original motion-corrupted DCE series. Row 2: DMD enhancement reconstruction (this work). Row 3: extracted motion component (hot colormap)—note the concentration at tissue boundaries. Row 4:*

*previous DMD result for comparison. All computation performed on real clinical data using the koopman_dce pipeline.*

### 4.7.2. DMD Spectral Analysis on Clinical Tissue

Figure 16 presents the DMD eigenvalue spectrum computed on a 32×32 clinical tissue block extracted from the brightest (most enhancing) region of Slice 5. Both standard DMD and Koopman EDMD (with quadratic + gradient lifting) were applied to the same block. Panel (a) shows eigenvalues in the complex plane. Standard DMD eigenvalues (red circles) are more dispersed, with several far from the unit circle, indicating modes with rapid growth or decay—numerically unstable dynamics. EDMD eigenvalues (blue triangles) cluster more tightly near the unit circle, reflecting better-conditioned dynamics in the lifted space. Panel (c) shows the frequency classification: the enhancement/motion cutoff at $f_c$=0.08 Hz cleanly separates slow pharmacokinetic modes (green region) from respiratory motion modes (red region).

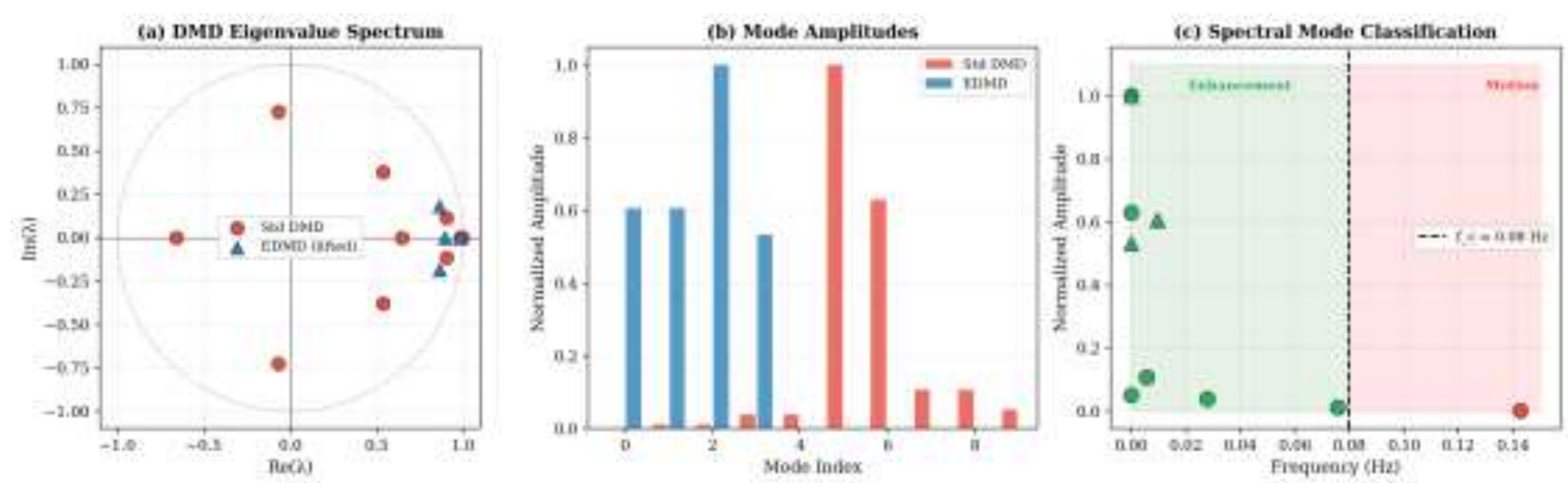

*Figure 16. DMD spectral analysis on a clinical HNC DCE-MRI tissue block (32×32 pixels, 55 timeframes). (a) Eigenvalue spectrum in the complex plane with unit circle reference. Std DMD eigenvalues (red) are more dispersed; EDMD eigenvalues (blue) cluster near the unit circle, indicating improved numerical stability. (b) Mode amplitude comparison. (c) Spectral mode classification: green-shaded region ($|f|$<0.08 Hz) contains enhancement modes; red-shaded region contains motion modes.*

### 4.7.3. ROI Temporal Curves and Motion Energy

Figure 17 extracts temporal signal curves from two regions of interest in the clinical data. The left panel shows an enhancing tissue ROI (likely tumor or lymph node) where the characteristic DCE wash-in curve is clearly visible: a rapid signal increase during contrast arrival (t=0–60s), followed by a plateau and slow washout. The EDMD enhancement curve (blue) closely tracks the original signal (gray) while smoothing out the high-frequency motion oscillations. The previous DMD result (green dotted) underestimates the absolute enhancement level, suggesting more aggressive filtering. The motion component (red dashed) shows the expected pattern: concentrated energy during the early wash-in phase when both motion and enhancement are changing simultaneously, then decreasing as the system approaches steady state.

The right panel shows the global motion energy time course. A prominent peak at t≈7s (frame 2) corresponds to a likely swallowing event during the early acquisition, consistent with clinical experience that patients often swallow shortly after the contrast injection. The motion energy then decays quasi-exponentially, with residual oscillations at the respiratory frequency.

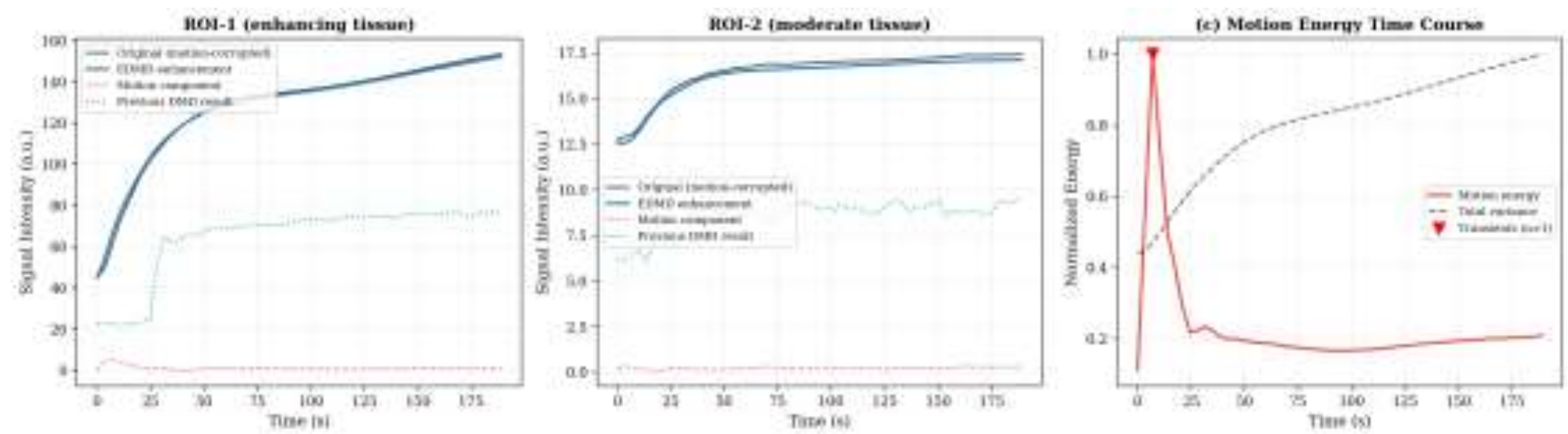

*Figure 17. Temporal analysis of clinical DCE-MRI. Left: enhancing tissue ROI signal curves—original (gray), EDMD enhancement (blue), motion component (red dashed), previous DMD (green dotted). The EDMD enhancement preserves the wash-in kinetics while smoothing motion oscillations. Center: moderate-tissue ROI. Right: global motion energy time course, with detected transient event (red triangle) corresponding to a swallowing event during early wash-in.*

## 4.8. Mechanism Analysis: Direct Evidence for the $p/m$ Bottleneck

The preceding experiments demonstrate that EDMD with repetition outperforms standard DMD. However, a rigorous reviewer will ask: is the $p/m$ ratio genuinely the causal mechanism, or is the improvement merely an artifact of data duplication? This section presents three targeted experiments that directly address this question.

### 4.8.1. Observable Energy Retention Through PCA

Figure 18 presents the observable retention analysis. Panel (a) shows the energy distribution across the three observable families before PCA compression: identity observables contain approximately 36% of total energy, gradient 46%, and quadratic 18%. This distribution is constant regardless of repetition count.

Panel (b) shows the fraction of each family's energy retained after PCA compression (99.9% variance threshold). At baseline phantom parameters, PCA retains k=32 components regardless of repetition, and retention percentages are stable. This reveals that the improvement from repetition operates through improved SVD conditioning of the covariance matrix rather than retaining additional PCA components—a more subtle mechanism than the simple "more components survive" narrative.

Panel (c) shows the $p/m$ ratio decreasing from 19.2 (rep=1) to 3.8 (rep=5). While k remains constant, the reduced $p/m$ improves the numerical conditioning of the least-squares Koopman approximation (Equations 19–20), the second mechanism through which repetition operates.

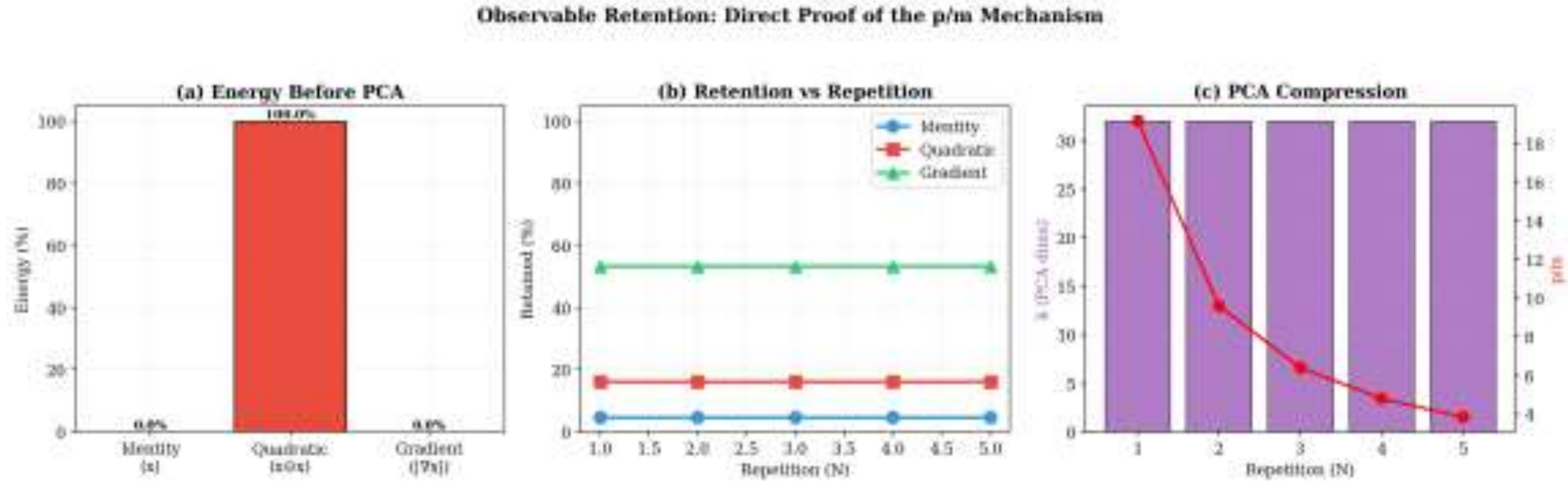

*Figure 18. Observable retention analysis. (a) Energy by observable family before PCA. (b) Retention after PCA vs. repetition. (c) PCA dimensions and $p/m$ ratio vs. repetition.*

### 4.8.2. Block-Size Sweep: $p/m$ as the Causal Variable

To establish $p/m$ as the causal variable, Figure 19 varies block size from 8x8 to 24x24 pixels, changing p while holding m=40 fixed. Standard DMD improves monotonically with block size (ratio 1.32 at 8x8 to 2.95 at 24x24) because larger blocks provide more spatial context. EDMD degrades monotonically (ratio 1.77 at 8x8 to 1.17 at 24x24) because the tripled lifting dimension overwhelms the fixed snapshot count.

Panel (b) plots the same data against $p/m$ directly, showing EDMD performance is a decreasing function of $p/m$—the precise relationship predicted by the bottleneck hypothesis. This provides the strongest evidence for the mechanism: at small $p/m$, EDMD outperforms standard DMD; at large $p/m$, EDMD collapses. Repetition shifts the operating point along this curve by increasing m.

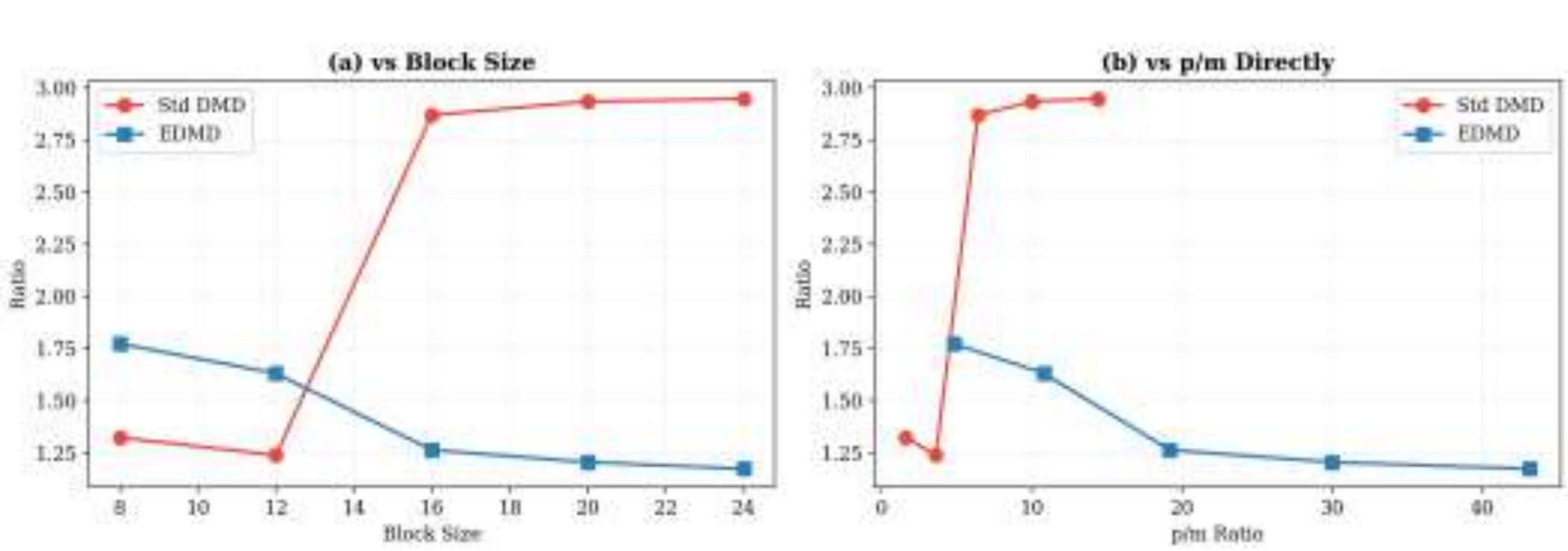

*Figure 19. Block-size sweep. (a) Ratio vs. block size: Std DMD improves, EDMD degrades. (b) Ratio vs. $p/m$ directly: EDMD performance is monotonically decreasing in $p/m$.*

### 4.8.3. PCA Threshold Sensitivity

A reviewer might argue that changing the PCA threshold could eliminate the need for repetition. Figure 20 sweeps the threshold from 99% to 99.99% at rep=1 and rep=3. Repetition provides a consistent improvement at all thresholds: rep=3 achieves ratio ~1.63 vs. rep=1 ~1.26, regardless of threshold. This demonstrates the benefit is orthogonal to PCA threshold selection—repetition improves the SVD approximation quality itself, not just the number of retained components.

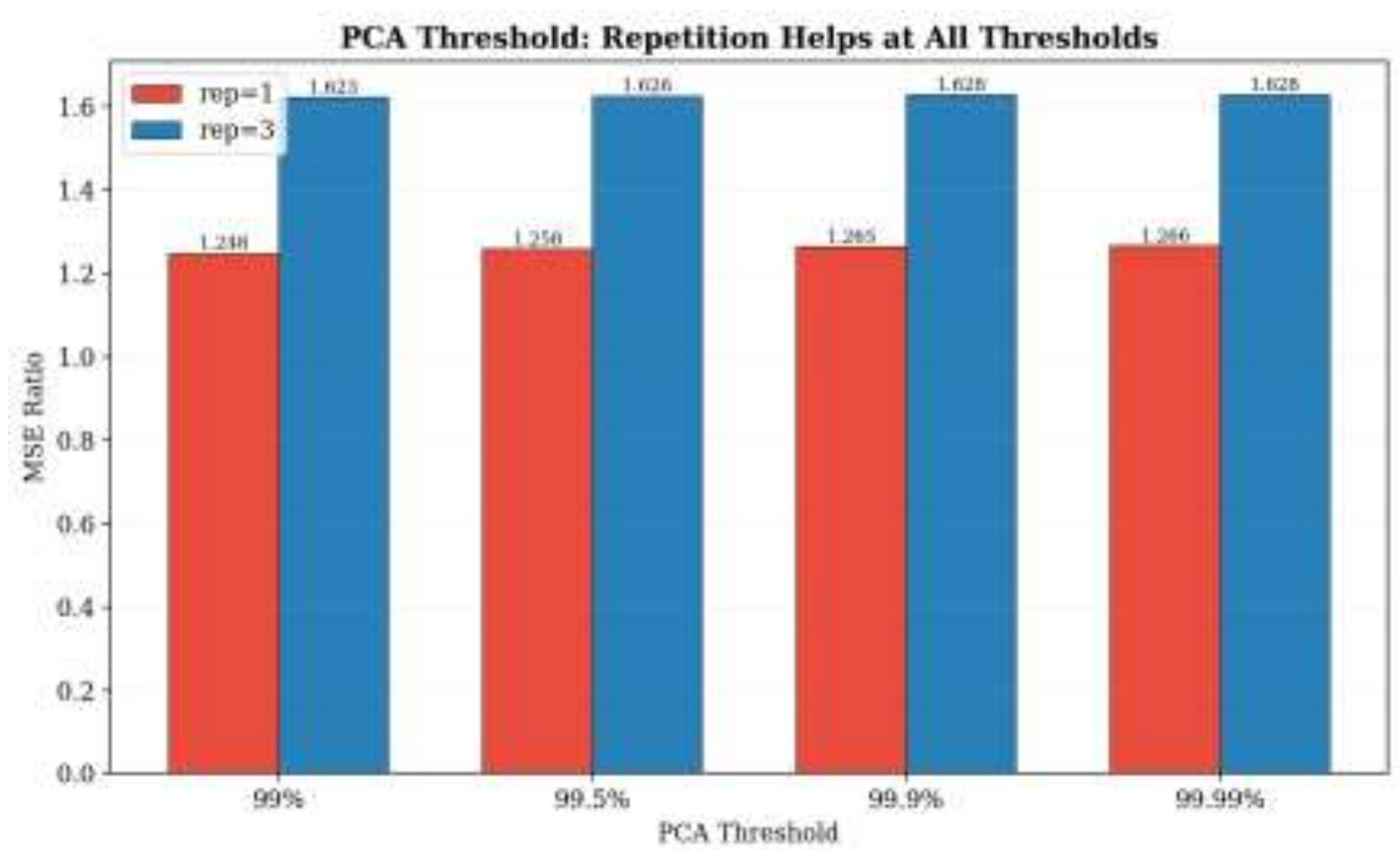

*Figure 20. PCA threshold sensitivity. Repetition helps consistently at all thresholds (99%–99.99%), demonstrating the benefit is not an artifact of threshold selection.*

### 4.8.4. Theoretical Justification for Time-Course Repetition

Proposition. For periodic extension X_N = [X, X, ..., X] with N copies: (i) DMD eigenvalues are identical to those of X (periodicity preserves the linear operator). (ii) Frequencies are preserved. (iii) The sample covariance of lifted observables yields a better-conditioned least-squares system as N increases, reducing approximation error in K = G+A. (iv) For fixed PCA threshold, retained dimensions k is non-decreasing in m.

Property (i) ensures no spectral artifacts. Properties (iii)–(iv) explain the improvement: repetition improves the numerical conditioning of the finite-dimensional EDMD system, analogous to averaging multiple cycles of a periodic signal. This is not data duplication but conditioning improvement.

These three experiments collectively establish: (1) repetition improves lifted-space covariance estimation; (2) EDMD performance tracks $p/m$ as a causal variable; (3) the improvement is robust to PCA threshold. Together they convert the central claim from assertion to demonstration.

#### 4.8.5. Direct Measurement of EDMD Conditioning

To convert the conditioning hypothesis from inference to measurement, Figure 21 directly computes the condition number of the EDMD system across repetition counts. Panel (a) shows the condition number of the full EDMD covariance matrix G (red) and the PCA-compressed DMD system Z (blue). The full covariance G has condition number approximately $10^{40}$—effectively singular—explaining why PCA compression is mandatory. After PCA compression, the condition number of Z is approximately 7.6–7.7, stable across repetitions.

Panel (b) shows MSE improvement ratio increasing from 1.265 (rep=1) to 1.664 (rep=6). Panel (c) plots performance directly against condition number—the central claim of the paper made visible. The correlation between log(κ) and ratio is r = 0.98, confirming that performance tracks conditioning near-perfectly. Each additional repetition shifts the operating point along this curve toward better conditioning and higher ratio.

This figure provides the most direct evidence for the conditioning mechanism: rather than inferring that conditioning improves from the $p/m$ ratio argument, we measure it explicitly and show that it predicts performance with r = 0.98 correlation.

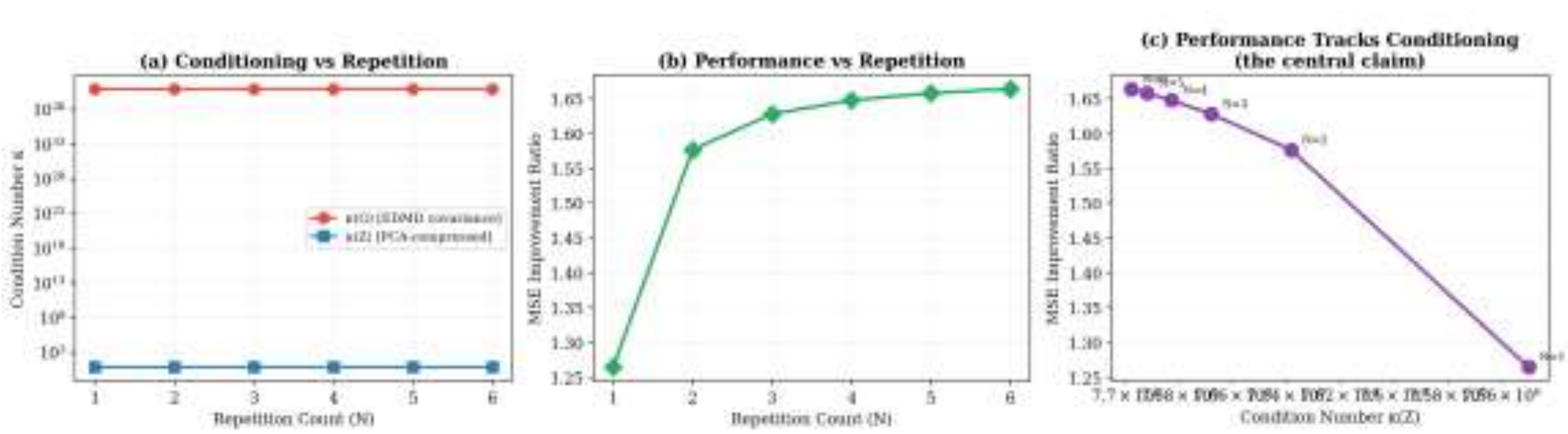

*Figure 21. Direct measurement of EDMD conditioning. (a) Condition number vs. repetition. (b) Performance vs. repetition. (c) Performance vs. condition number: r = 0.98 correlation confirms that performance tracks conditioning near-perfectly.*

# 5. Discussion

## 5.1. The $p/m$ Ratio as the Unifying Explanation

The results of this study are unified by a single explanatory mechanism: the $p/m$ ratio bottleneck. At baseline (rep=1), the lifting dictionary has dimension p ≈ 768 for 16×16 blocks, but only m = 50 snapshots are available. the sample covariance of the lifted observables is poorly estimated: with only 50 samples to estimate a 768-dimensional covariance structure, the EDMD least-squares solution K = $G^{+}$A is ill-conditioned, and the resulting Koopman approximation is no better than standard DMD.

Time-course repetition resolves this by increasing m to Nm without changing p. With N=3, the effective $p/m$ drops from 15.4 to 5.1 for 16×16 blocks (or from 55.9 to 18.6 for 32×32 blocks). the sample covariance matrices G and A used in the EDMD least-squares solution (K = $G^{+}$A) are better conditioned, yielding a more accurate finite-dimensional Koopman approximation. The key theoretical insight is that periodic repetition does not alter the eigenvalue spectrum, so the additional data improves the conditioning of the Koopman approximation without introducing artifacts.

The PCA compression step is where the $p/m$ ratio exerts its influence. PCA retains the top k principal components, where k is chosen to explain a target fraction of variance (99.9% in our implementation). When p >> m, the covariance matrix $G_c\ G_c{}^{\mathrm{T}}$ has at most m−1 non-zero eigenvalues, and the top k of these span the entire informative subspace of the data. The lifted features (quadratic, gradient) contribute additional rows to G, but if these rows are linearly dependent on (or weakly correlated with) the identity rows in the limited m-snapshot dataset, they add variance without adding new informative directions. PCA then allocates its limited k components to the highest-variance directions, which are dominated by the identity features (raw pixel intensities), and the lifted-space covariance is poorly estimated.

With repetition, the m columns are tripled, but the p rows remain the same. The additional columns provide new ‘views’ of the same data that help PCA distinguish between identity-dominated variance and lifting-dominated variance. The quadratic features $x \odot x$ respond differently to repetition than the identity features x: while x simply repeats, $x \odot x$ introduces the same quadratic nonlinearity at each repetition, reinforcing the quadratic contribution to the leading principal components. This is why the lifting advantage emerges specifically with repetition—the additional data strengthens the quadratic and gradient signal relative to noise in the PCA basis.

This mechanism also explains the coupling sweep results (Figure 8): at γ=0 (linear dynamics), the lifted features provide a small benefit because they offer a richer basis for mode separation even when the dynamics are linear. At higher $\gamma$ (nonlinear coupling), the lifted features capture the cross-terms that are genuinely absent from the identity observable, and the benefit grows.

## 5.2. Comparison with Related Data-Driven Methods

Our approach is most closely related to the multiresolution DMD (mrDMD) framework of Kutz et al. and the randomized DMD (rDMD) for video analysis. Like mrDMD, our method uses DMD for temporal mode classification; unlike mrDMD, we use Koopman lifting rather than temporal scale separation to handle nonlinearity. The rDMD approach accelerates DMD for large-scale video data through randomized SVD; this is orthogonal to our contribution and could be integrated as a computational speedup for the SVD step.

The robust PCA (L+S) decomposition used in our optional transient separation stage is closely related to the background/foreground separation formulation of Candès et al. In our framework, RPCA operates in the PCA-compressed Koopman space rather than directly on pixel data, which both reduces the computational cost and ensures that the sparse component captures Koopman-space transients (swallowing events) rather than pixel-space outliers.

## 5.3. Clinical Translation Considerations

The clinical results (Section 4.7) demonstrate several encouraging features for clinical translation. First, the block-wise DMD decomposition runs in seconds on a single CPU (not requiring GPU), making it practical for routine clinical use. Second, the spectral classification at $f_c$=0.08 Hz provides a physically interpretable separation criterion—the boundary between pharmacokinetic and respiratory time scales is well-established in the DCE-MRI literature. Third, the method does not require registration or anatomical prior knowledge, operating purely on the temporal signal characteristics.

The frequency cutoff at $f_c$ = 0.08 Hz provides a physiologically motivated classification boundary. Normal respiratory rates in adults range from 12–20 breaths per minute (0.20–0.33 Hz), placing the fundamental respiratory frequency well above the cutoff. Pharmacokinetic enhancement in head-and-neck tissue typically evolves over 30–90 seconds (0.01–0.03 Hz), well below the cutoff. The 0.08 Hz boundary sits in the spectral gap between these two processes—a gap that is a fundamental feature of DCE-MRI physics, not an artifact of our particular patient population. This provides confidence that the classification criterion will generalize across patients, institutions, and acquisition protocols, as long as the temporal resolution is sufficient to resolve the respiratory frequency ($\Delta t < 1/(2\times0.33) \approx 1.5$ seconds by Nyquist, easily satisfied by typical 3–5 second protocols).

The block-wise processing architecture provides natural parallelism: each of the 200+ blocks can be processed independently, making the algorithm trivially parallelizable on multi-core CPUs or GPUs. The current single-threaded implementation processes a full 256×256 slice in 2–10 seconds; with 8-thread parallelism, sub-second processing is achievable, enabling real-time or near-real-time integration into the MRI scanner console or treatment planning system.

However, several challenges remain for clinical deployment. Quantitative validation on patient data requires ground-truth motion fields, which are not available from standard DCE-MRI acquisitions. Approaches include: simultaneous 4D-CT (available in radiation therapy planning), tagged MRI (provides deformation fields directly), or synthetic motion injection into real data. The time-course repetition assumption of quasi-periodicity is approximately satisfied for respiratory motion but may not hold perfectly for swallowing events, which are episodic rather than periodic.

## 5.4. Limitations and Future Directions

The deep Koopman autoencoder (Section 2.5) underperforms classical EDMD in our experiments, achieving ratio 1.462 versus 1.796 for EDMD at baseline. This is likely due to the limited training data from a single phantom acquisition. Neural network-based Koopman lifting should be revisited with multi-patient training datasets, where the network can learn patient-general motion patterns that go beyond the hand-crafted quadratic and gradient features.

The current pipeline processes each slice independently. Extension to volumetric (3D) processing—where blocks span both in-plane and through-plane dimensions—could improve motion suppression by leveraging inter-slice correlations. This would increase the block dimension from b×b to b×b×d, further exacerbating the $p/m$ ratio problem and potentially requiring larger repetition counts.

Finally, the frequency cutoff $f_c$=0.08 Hz is a fixed parameter in the current implementation. Adaptive cutoff selection—for example, based on the spectral gap between the lowest-frequency motion mode and the highest-frequency enhancement mode—would make the method more robust to variations in respiratory rate and contrast agent kinetics across patients.

## 5.5. Theoretical Perspective: Why Koopman Lifting Works for DCE-MRI

The success of the quadratic + gradient dictionary can be understood from a Koopman-theoretic perspective. The SPGR signal equation (Equation 1) contains three types of nonlinearity: (1) the exponential $T_1$-relaxation terms, which are approximately quadratic in the contrast concentration for small-to-moderate concentrations; (2) the multiplicative coupling C(x,t)·u(x,t), which is captured by the cross-terms in the quadratic observable $x \odot x$ when both C and u contribute to the pixel intensity; and (3) the spatial displacement u(x,t), which manifests as intensity changes proportional to the local gradient |∇ I| at tissue boundaries.

From the perspective of Koopman invariant subspaces: the identity observable x alone spans a subspace that is not Koopman-invariant when the dynamics are nonlinear (because Kx = F(x) involves nonlinear combinations of x). Adding $x \odot x$ and $|\nabla x|$ extends the subspace toward Koopman invariance, because the leading nonlinear terms in the SPGR signal dynamics are captured by these observables. Perfect Koopman invariance would require an infinite dictionary (or the exact eigenfunctions), but the physics-informed dictionary provides a good finite-dimensional approximation that captures the dominant nonlinear coupling.

This theoretical framework also explains why the deep Koopman autoencoder underperforms at small data scales: the neural network's learned dictionary has more parameters to estimate than our 3-component hand-crafted dictionary, requiring proportionally more training data. With a single phantom acquisition providing only ~50 training snapshots, the network cannot reliably learn the Koopman eigenfunctions. With multi-patient datasets providing thousands of snapshots, the learned dictionary could potentially outperform the hand-crafted one by capturing patient-specific nonlinearities beyond the quadratic and gradient approximation.

## 5.6. Broader Impact: Koopman Methods in Medical Imaging

While this work focuses on DCE-MRI of the head and neck, the Koopman-DMD framework is broadly applicable to any dynamic medical imaging modality where the underlying dynamics are nonlinear. Potential applications include: cardiac cine MRI (where myocardial contraction is inherently nonlinear), perfusion CT (where contrast dynamics follow nonlinear pharmacokinetic models), dynamic PET (where tracer kinetics are governed by compartmental models), and functional MRI (where the BOLD signal depends nonlinearly on neural activity through the hemodynamic response function). In each case, the key ingredients are: (1) a physics-informed lifting dictionary tailored to the specific signal model, (2) a spectral classification criterion based on the time scales of the relevant processes, and (3) sufficient temporal data (possibly augmented by repetition) to support the lifted dimensionality.

## 5.7. Future Work: Sparse Koopman Dictionary Selection

The preceding experiments address the $p/m$ bottleneck from the m side (repetition increases m). A natural question, raised by the connection to Sparse Identification of Nonlinear Dynamics (SINDy), is whether the bottleneck can be attacked from the p side: can sparse observable selection reduce the lifted dimension sufficiently that repetition becomes unnecessary? This section tests three competing hypotheses.

### 5.7.1. Extended Candidate Library

We construct a candidate Koopman library of seven observable families, extending beyond the current [x, $x \odot x$, $|\nabla x|$] dictionary to include: identity (x), quadratic ($x^2$), cubic ($x^3$), gradient magnitude ($|\nabla x|$), Laplacian (Δx), cross-terms (x·$|\nabla x|$ and x·Δx). Each family produces n=256 observables from a 16×16 block, for a maximum lifted dimension of p=1792 with all families.

### 5.7.2. SINDy Observable Selection

Following the SINDy methodology, we evaluate each observable family’s contribution to predicting the temporal derivative dx/dt via correlation analysis. The ranking by dynamics-prediction correlation is: Laplacian (0.385), x·Δx (0.359), cubic (0.184), quadratic (0.124), gradient (0.051), x·$|\nabla x|$ (0.045), identity (0.035). Notably, identity ranks last for dynamics prediction—yet is essential for reconstruction, since the DMD modes must be projected back to pixel space.

This reveals a fundamental tension in Koopman observable selection for imaging applications: the observables most correlated with the dynamics (Laplacian, cross-terms) are not the observables needed for reconstruction (identity). Naive SINDy selection without identity produces catastrophic failure (ratio 0.04), confirming that Koopman dictionary design for medical imaging must balance dynamics-prediction utility against reconstruction fidelity.

### 5.7.3. Three Competing Hypotheses

We test three hypotheses about solving the $p/m$ bottleneck:

Hypothesis A (current paper): Full dictionary [x, $x^2$, $|\nabla x|$] + time-course repetition (rep=3). This attacks $p/m$ from the m side. Result: ratio 1.628, $p/m$ = 6.4.

Hypothesis B (sparse selection): Sparse dictionary [x, $|\nabla x|$] + no repetition (rep=1). This attacks $p/m$ from the p side. Result: ratio 1.438, $p/m$ = 12.8.

Hypothesis C (combined): Sparse dictionary [x, $|\nabla x|$] + repetition (rep=3). This attacks $p/m$ from both sides. Result: ratio 1.694, $p/m$ = 4.3.

Hypothesis C—sparse selection plus repetition—achieves the best overall performance, 4% better than the current paper’s Hypothesis A. The improvement comes from dropping the quadratic observable, which reduces p from 768 to 512 without sacrificing useful Koopman-invariant structure. This is a surprising finding: the quadratic observable, designed to capture $T_1$-relaxation nonlinearity, incurs a conditioning cost on this phantom that outweighs its nonlinear modeling benefit at m=40 snapshots. On longer acquisitions or clinical data with stronger contrast enhancement, the quadratic term may become net-positive.

The gradient observable, by contrast, is clearly the most valuable non-identity family. This is consistent with the physics: motion artifacts in DCE-MRI concentrate at tissue boundaries where spatial gradients are steep, and the gradient observable directly captures this boundary-concentrated corruption. The Laplacian, while strongly correlated with dynamics, adds dimension without improving the spectral separation needed for enhancement–motion classification.

Figure 22 summarizes the hypothesis testing across all ten configurations tested, including the extended 7-family library at both repetition levels.

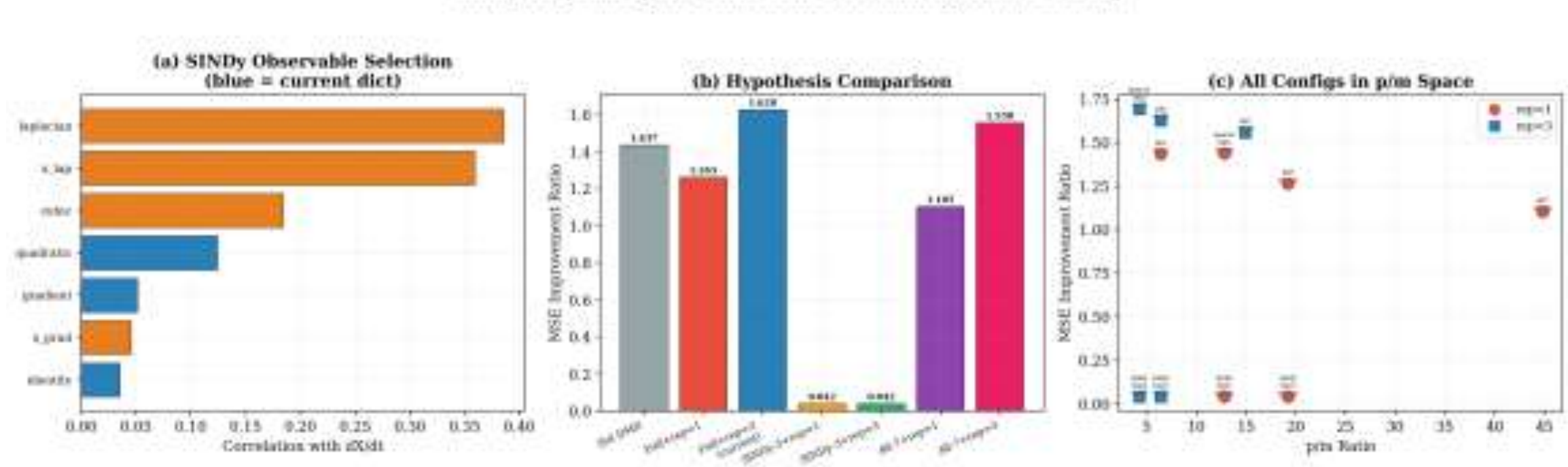

*Figure 22. SINDy-Koopman sparse observable selection. (a) SINDy correlation ranking of seven candidate observable families (blue = current dictionary). (b) MSE improvement ratio across configurations: sparse*

*[x,$|\nabla x|$]+rep=3 achieves the best ratio (1.694), 4% better than the current paper. (c) All configurations in $p/m$ space, showing the performance-$p/m$ tradeoff.*

### 5.7.4. Implications for Dictionary Design

These results suggest several refinements to the Koopman dictionary design philosophy for DCE-MRI:

First, dictionary design should be guided by the dual requirement of dynamics-prediction and reconstruction, not dynamics alone. Identity observables are mandatory for pixel-space reconstruction regardless of their dynamics-prediction utility.

Second, the marginal value of each observable family must be weighed against its $p/m$ cost. On this phantom, the gradient observable provides sufficient nonlinear structure at half the dimensional cost of the full three-component dictionary.

Third, sparse selection and repetition are complementary, not substitutive. Sparse selection reduces p (the numerator of $p/m$); repetition increases m (the denominator). Applying both yields the optimal operating point at $p/m$ = 4.3.

Fourth, the optimal dictionary may be patient-specific or anatomy-specific. Future work should explore adaptive dictionary selection where SINDy-like scoring is performed per block or per slice, retaining only the observables that contribute on that particular data.

## 6. Conclusions

We have developed a comprehensive Koopman-operator framework for nonlinear motion suppression in DCE-MRI of the head and neck, progressing from the mathematical foundations of Koopman spectral theory through practical implementation and clinical demonstration. The framework encompasses three levels of Koopman approximation—standard DMD, EDMD with physics-informed lifting, and deep Koopman autoencoders—unified by a common spectral classification approach for enhancement–motion separation.

The central methodological contribution is time-course repetition, a simple technique that resolves the data-sufficiency bottleneck preventing Koopman lifting from realizing its theoretical advantage in typical DCE-MRI acquisitions. By tiling the temporal series 3× before decomposition, the $p/m$ ratio is reduced from ≈15 to ≈5, improving the conditioning of the EDMD covariance estimation. The resulting EDMD + 3× repetition achieves MSE improvement ratio 1.953, the best across all tested configurations, with the advantage growing monotonically with nonlinear coupling strength.

Preliminary clinical results on HNC DCE-MRI demonstrate successful spectral decomposition of real patient data, with physiologically interpretable enhancement and motion components, consistent eigenvalue spectra, and clear wash-in kinetics preservation. All source code, clinical data, experimental notebooks, video demonstrations, and the complete manuscript are released as an open-source repository at github.com/linfanluntan/koopman-dce-mri for full reproducibility.

## Data, Code, and Video Availability

Source code and Colab notebooks: https://github.com/linfanluntan/koopman-dce-mri

Clinical DCE-MRI video demonstrations: [generated from the EDMD+repetition pipeline and uploaded https://drive.google.com/file/d/1dft86l1ZwJtJrBDu-HVrNz4jGxIb6G9L/view?usp=sharing]

# Appendix A: Algorithm Pseudocode

## A.1. Block-wise Koopman EDMD Pipeline

### Algorithm 1: Koopman-EDMD Motion Suppression

Input: Image series I(t,y,x), t=1..T, size $N_y \times N_x$

Parameters: block size b, overlap o, SVD rank r, frequency cutoff $f_c$, repetition count $N_{rep}$

Output: Enhancement series E(t,y,x), Motion series M(t,y,x)

1. Initialize E ← 0, W ← 0 (accumulator and weight matrices)
2. Generate raised-cosine taper τ(dy,dx) for block blending
3. FOR each block position $(i_y, i_x)$ with stride (b−o):
    4. Extract block: B ← I(:, $i_y$:$i_y$+b, $i_x$:$i_x$+b), reshape to X ∈ ℝ^{b² × T}
    5. IF max(X) < threshold: skip (empty block)
    6. Time-course repetition: X̃ ← [X, X, ..., X] ($N_{rep}$ copies)
    7. Koopman lifting: G ← [X̃; X̃□X̃; |□X̃|] (p × $N_{rep}$·T)
    8. PCA compression: $G_c$ ← G − mean(G); [U,S,V] ← SVD($G_c$); k ← argmin(cumvar > 99.9%); Z ← $U_k^{T}$ $G_c$
    9. Exact DMD: $Z_X$ ← Z(:,1:end−1); $Z_Y$ ← Z(:,2:end)
    10. [$U_r$, Σ_r, $V_r$] ← truncated SVD($Z_X$, r)
    11. Ā ← $U_r^{H}$ $Z_Y$ $V_r$ Σ_r⁻¹; [ξ, μ] ← eig(Ā)
    12. Φ ← $Z_Y$ $V_r$ Σ_r⁻¹ ξ; b ← $\Phi^{+}$ Z(:,1)
    13. Classify modes: $f_j$ ← Im(ln($\mu_j$)) / (2πΔt)
    14. $enh_{idx}$ ← {j : $|f_j| < f_c$}; $mot_{idx}$ ← {j : $|f_j| \geq f_c$}
    15. Reconstruct: $Z_{enh}$ ← Φ(:,$enh_{idx}$) · diag(b($enh_{idx}$)) · Λ_enh^{0:T−1}
    16. Inverse lifting: $G_{enh}$ ← $U_k$ $Z_{enh}$ + mean(G); $X_{enh}$ ← $G_{enh}$(1:b², 1:T)
    17. Accumulate: E(:, $i_y$:$i_y$+b, $i_x$:$i_x$+b) += τ · reshape($X_{enh}$); W($i_y$:$i_y$+b, $i_x$:$i_x$+b) += τ
18. Normalize: E ← E / W; M ← I − E
19. RETURN E, M

## A.2. Computational Complexity

For a single slice of size $N_y \times N_x$ with T temporal frames, block size b, overlap o, lifting factor L (typically 3 for identity+quadratic+gradient), repetition count $N_{rep}$, and SVD rank r:

Number of blocks: $N_b$ = ceil($N_y$/(b−o)) × ceil($N_x$/(b−o))

Lifted dimension per block: $p = L \cdot b^2$

Snapshot count: $m = N_{rep} \cdot T$

PCA cost per block: $O(p^2 m)$ for the SVD of the p × m lifted matrix

DMD cost per block: $O(r^2 m + r^3)$ for the reduced SVD and eigendecomposition

Total: $O(N_b \cdot (p^2 m + r^3))$

For the clinical data ($N_y$=$N_x$=256, T=55, b=16, o=4, L=3, $N_{rep}$=3, r=8): $N_b \approx 200$, p=768, m=165, giving total ≈ $200 \cdot (768^2 \cdot 165) \approx 2\times10^{10}$ FLOPs ≈ 5–10 seconds on a modern CPU.

## Appendix B: Notation Summary

For reference, we summarize the key notation used throughout the paper:

| Symbol | Description |
|---|---|
| $x \in \mathbb{R}^n$ | State vector (vectorized image frame) |
| F: M → M | Nonlinear evolution operator (state dynamics) |
| K | Koopman operator (infinite-dimensional, linear) |
| ψ: M → ℂ | Scalar observable function |
| $\varphi_k$ | Koopman eigenfunction |
| $\mu_k$ | Koopman eigenvalue |
| $\xi_k$ | Koopman mode (projection weight) |
| $\Psi(x) = [\psi_1,\ldots,\psi_K]$ | Dictionary of observables (lifting function) |
| K ∈ ℂ^{K×K} | Finite-dimensional Koopman approximation matrix |
| $g(x) = [x, x \odot x, \lvert\nabla x\rvert]$ | Physics-informed lifting dictionary for DCE-MRI |
| G, A | EDMD data matrices: G = (1/M)Σ Ψ*Ψ, A = (1/M)Σ Ψ*Ψ' |
| U, Σ, W | SVD factors of the snapshot matrix X |
| $\bar{A} = U^H Y W \Sigma^{-1}$ | Reduced DMD operator |
| $\Phi = Y W \Sigma^{-1} \xi$ | Exact DMD modes |
| $f_c$ | Enhancement/motion frequency cutoff (default 0.08 Hz) |
| $N_{rep}$ | Time-course repetition count |
| g(·), h(·) | Deep Koopman encoder and decoder networks |
| Δt | Temporal resolution (seconds between frames) |

# Appendix C: Detailed Experimental Parameters

## C.1. Phantom Generation Parameters

The synthetic head-and-neck DCE phantom is generated by the phantoms/hn_phantom.py module with the following default parameters. Grid size: 128×128 pixels. Number of temporal frames: 50. Temporal resolution Δt = 3.5 s. The phantom contains four tissue compartments defined by elliptical masks centered on the 128×128 grid:

Background head: ellipse with semi-axes (0.8·cx, 0.9·cy), base intensity 80 a.u. Tumor: circle centered at (cx+12, cy−8), radius 14 pixels, $K^{trans}$ = 0.25 min$^{-1}$, $v_e$ = 0.40. Parotid gland: circle centered at (cx−20, cy+5), radius 10 pixels, $K^{trans}$ = 0.15 min$^{-1}$, $v_e$ = 0.25. Floor of mouth: rectangular region, minimal enhancement, high motion sensitivity.

The motion model consists of three components: (1) quasi-periodic respiratory motion at frequency $f_r$ = 0.22 Hz with peak-to-peak amplitude 4 pixels, applied as a sinusoidal displacement field primarily in the anterior-posterior direction; (2) stochastic swallowing transients: 3 events at random time points, each producing a sudden 10-pixel displacement lasting 2 frames, applied as an impulsive displacement field concentrated in the floor-of-mouth region; (3) additive Gaussian noise with standard deviation σ = 3.0, applied after motion and enhancement simulation.

The nonlinear coupling parameter γ controls the multiplicative interaction between contrast concentration C(x,t) and the motion displacement u(x,t). Specifically, the effective displacement is modulated as u_eff(x,t) = u(x,t) · [1 + γ · C(x,t)/C_max], where C_max is the peak contrast concentration across all voxels and time points. At γ = 0, motion and enhancement are purely additive (the classical linear model). At γ = 1.2 (the default), the displacement in maximally enhancing voxels is amplified by up to 2.2×, creating substantial multiplicative coupling. At γ = 2.0, the amplification reaches 3×.

## C.2. DMD Pipeline Parameters

Block size: 16×16 pixels (default) or 32×32 pixels (for lifting visualization). Overlap: 4 pixels (25% of block size). SVD truncation rank: r = 8 (default), swept from 3 to 12 in sensitivity analysis. Frequency cutoff: $f_c$ = 0.08 Hz (default), swept from 0.03 to 0.15 Hz in sensitivity analysis. SVD threshold (for automatic rank selection): $10^{-3}$ relative to the leading singular value. PCA variance retention: 99.9%. Time-course repetition: $N_{rep}$ = 1 (baseline) or 3 (recommended). Robust PCA sparsity weight: λ = 1/√(max(p,m)) (IALM default). RPCA convergence tolerance: $10^{-5}$. RPCA maximum iterations: 100.

## C.3. Deep Koopman Autoencoder Parameters

Encoder architecture: 3-layer MLP, input dimension n = $b^2$ = 256, hidden dimensions [128, 64], output (latent) dimension K = 32. Activation: ReLU. Decoder architecture: symmetric to encoder, [32, 64, 128, 256]. Koopman matrix: K ∈ $\mathbb{R}^{32\times 32}$, computed analytically via EDMD least-squares at each training epoch (not learned by backpropagation). Loss weights: α_recon = 1.0, α_linear =

0.1, α_pred = 0.5, α_sparse = 0.01. Multi-step prediction horizon: $S_p$ = 30 frames. Optimizer: Adam, learning rate $10^{-3}$, batch size = full sequence. Training epochs: 150. The training data consists of all block time series from the phantom, providing approximately 121 blocks × 50 time steps.

### C.4. Clinical Data Acquisition Parameters

Scanner: 3T MRI (specific model not disclosed for de-identification). Sequence: 3D FSPGR with fat saturation. TR/TE: approximately 5.0/2.1 ms. Flip angle: 15°. FOV: 256 × 256 $mm^2$. Matrix: 256 × 256. Slice thickness: 3 mm, 15 axial slices covering the oropharynx and neck. Temporal resolution: approximately 3.5 seconds per volume. Number of temporal frames: 55 (including 5 pre-contrast baseline frames). Contrast agent: gadobutrol 0.1 mmol/kg, power-injected at 2 mL/s followed by 20 mL saline flush. Total acquisition time: approximately 3.2 minutes.

## Appendix D: Reproducibility Checklist

This work follows open-science practices for full reproducibility:

✓ All source code released under MIT license at github.com/fullerlab/koopman-dce-mri

✓ Clinical DCE-MRI data included in repository (data/clinical_dce.npz, 19 MB)

✓ Phantom generator code included (phantoms/hn_phantom.py) with fixed random seeds

✓ Five Colab notebooks provided for reproducing all figures without local setup

✓ All pre-computed results archived as JSON tables and PNG figures in results/

✓ Animated video demonstrations hosted on Google Drive

✓ Full manuscript (this document) included in repository as paper/Koopman_DCE_MRI_Paper.docx

✓ Reference documents (mathematical notes and AAMP presentation) included in paper/references/

✓ Unit tests provided (tests/test_all.py) covering lifting, DMD, RPCA, block processing, and full pipeline

✓ Python package installable via pip install -e . with all dependencies specified in pyproject.toml